\documentclass[12pt]{article}

\usepackage{graphicx}
\usepackage{scicite}
\usepackage{times}
\usepackage[T1]{fontenc}
\usepackage[colorlinks,bookmarks=false,citecolor=magenta,linkcolor=magenta,urlcolor=magenta]{hyperref}

\usepackage{braket}
\usepackage{ulem}
\usepackage{epsfig}
\usepackage{epstopdf}
\usepackage{color}
\usepackage{float}

\newenvironment{sciabstract}{%
	\begin{quote} \bf}
	{\end{quote}}

\newcommand{\red}[1]{{\color{black} #1}}

\title{Quantum walks on a programmable two-dimensional 62-qubit superconducting processor}

\author
    {Ming Gong,$^{1,2,3\dag}$ Shiyu Wang,$^{1,2,3\dag}$ Chen Zha,$^{1,2,3\dag}$ Ming-Cheng Chen,$^{1,2,3}$\\
	He-Liang Huang,$^{1,2,3}$ Yulin Wu,$^{1,2,3}$ Qingling Zhu,$^{1,2,3}$ Youwei Zhao,$^{1,2,3}$\\
	Shaowei Li,$^{1,2,3}$ Shaojun Guo,$^{1,2,3}$ Haoran Qian,$^{1,2,3}$ Yangsen Ye,$^{1,2,3}$ \\
	Fusheng Chen,$^{1,2,3}$ Chong Ying,$^{1,2,3}$ Jiale Yu,$^{1,2,3}$ Daojin Fan,$^{1,2,3}$ \\
	Dachao Wu,$^{1,2,3}$ Hong Su,$^{1,2,3}$ Hui Deng,$^{1,2,3}$ Hao Rong,$^{1,2,3}$ Kaili Zhang,$^{1,2,3}$ \\
	Sirui Cao,$^{1,2,3}$ Jin Lin,$^{1,2,3}$ Yu Xu,$^{1,2,3}$ Lihua Sun,$^{1,2,3}$ Cheng Guo,$^{1,2,3}$ \\
	Na Li,$^{1,2,3}$ Futian Liang,$^{1,2,3}$ V. M. Bastidas,$^{4}$ Kae Nemoto,$^{5}$ W. J. Munro,$^{4,5}$\\
	Yong-Heng Huo,$^{1,2,3}$ Chao-Yang Lu,$^{1,2,3}$ Cheng-Zhi Peng,$^{1,2,3}$ Xiaobo Zhu,$^{1,2,3\ast}$ \\
	Jian-Wei Pan$^{1,2,3\ast}$\\
	\\
	\normalsize{$^1$  Hefei National Laboratory for Physical Sciences at the Microscale and Department}\\
	\normalsize{of Modern Physics, University of Science and Technology of China,}\\
	\normalsize{Hefei 230026, China}\\
	\normalsize{$^2$  Shanghai Branch, CAS Center for Excellence in Quantum Information and }\\
	\normalsize{Quantum Physics, University of Science and Technology of China,}\\
	\normalsize{Shanghai 201315, China}\\
	\normalsize{$^3$  Shanghai Research Center for Quantum Sciences, Shanghai 201315, China}\\
	\normalsize{$^4$  NTT Basic Research Laboratories and Research Center for Theoretical Quantum} \\
	\normalsize{Physics, 3-1 Morinosato-Wakamiya, Atsugi, Kanagawa 243-0198, Japan}\\
	\normalsize{$^5$  National Institute of Informatics, 2-1-2 Hitotsubashi,} \\
	\normalsize{Chiyoda-ku, Tokyo 101-8430, Japan}\\
	\\
	\normalsize{$^\ast$Corresponding authors. E-mail: xbzhu16@ustc.edu.cn (X.Z.); pan@ustc.edu.cn (J.-W.P.)}\\
	\normalsize{$\dag$ These authors contributed equally to this work.}
}

\begin{document}
	
	\baselineskip24pt
	
	\maketitle	
	\newpage
	\begin{sciabstract}
		Quantum walks are the quantum mechanical analogue of classical random walks and an extremely powerful tool in quantum simulations, quantum search algorithms, and even for universal quantum computing. In our work, we have designed and fabricated an 8x8 two-dimensional square superconducting qubit array composed of 62 functional qubits. We used this device to demonstrate high fidelity single and two particle quantum walks. Furthermore, with the high programmability of the quantum processor, we implemented a Mach-Zehnder interferometer where the quantum walker coherently traverses in two paths before interfering and exiting. By tuning the disorders on the evolution paths, we observed interference fringes with single and double walkers. Our work is an essential milestone in the field, brings future larger scale quantum applications closer to realization on these noisy intermediate-scale quantum processors. 
	\end{sciabstract}	
	
	The quantum principles of superposition and entanglement allow a more powerful form of random walk -- termed quantum walk (QW)~\cite{aharonov1993quantum}. 
	These walks have attracted considerable attention with many applications known in quantum transport~\cite{mulken2011continuous}, quantum simulation~\cite{aspuru2012photonic,lambert2013quantum}, quantum search algorithms~\cite{shenvi2003quantum,childs2004spatial}, and even universal quantum computing~\cite{childs2009universal,childs2013universal}. 
	\red{The universality of quantum walks has been shown with encoded quantum computation,
	meaning they are a universal quantum computation primitive and may give an exponential algorithmic speed-up~\cite{childs2003exponential}.} 
	\red{In particular, quantum walks with multiple walkers show a quantum advantage~\cite{rohde2012error,zhong2020quantum} superior to ones using only a single walker.} 
	Motivated by {QWs'} rich potential applications, numerous proof of principle experimental demonstrations have been performed in a wide variety of hardware platforms, \red{ranging from} photonics\cite{peruzzo2010quantum,chen2018observation}, trapped ions~\cite{schmitz2009quantum,zahringer2010realization}, neutral atoms~\cite{karski2009quantum}, \red{to} nuclear magnetic resonance~\cite{ryan2005experimental}, and superconducting qubits~\cite{ramasesh2017direct,yan2019strongly}.
	
	\red{It is well-known~\cite{tulsi2008faster} {that} QW based quantum search algorithms {require at} least a two-dimensional configuration.} 
	Furthermore, easy circuit programmability \red{with an arbitrary number of walkers is an essential requirement to encode applications,} where the configuration can be changed on a walk-by-walk basis including the adjustability of tunneling amplitude and graph structure~\cite{underwood2012bose}. Achieving both of these simultaneously have proved experimentally challenging. Superconducting circuits \red{provide the nonlinear interaction Hamiltonian necessary for the universal quantum computation, making them} one of the leading quantum computer approaches~\cite{arute2019quantum}. 
	Along with the real time programmability, they are now an excellent candidate system for the realization of fully configurable two-dimensional QWs.

\begin{figure}[H]
		\centering 
		\includegraphics[width=0.6\textwidth]{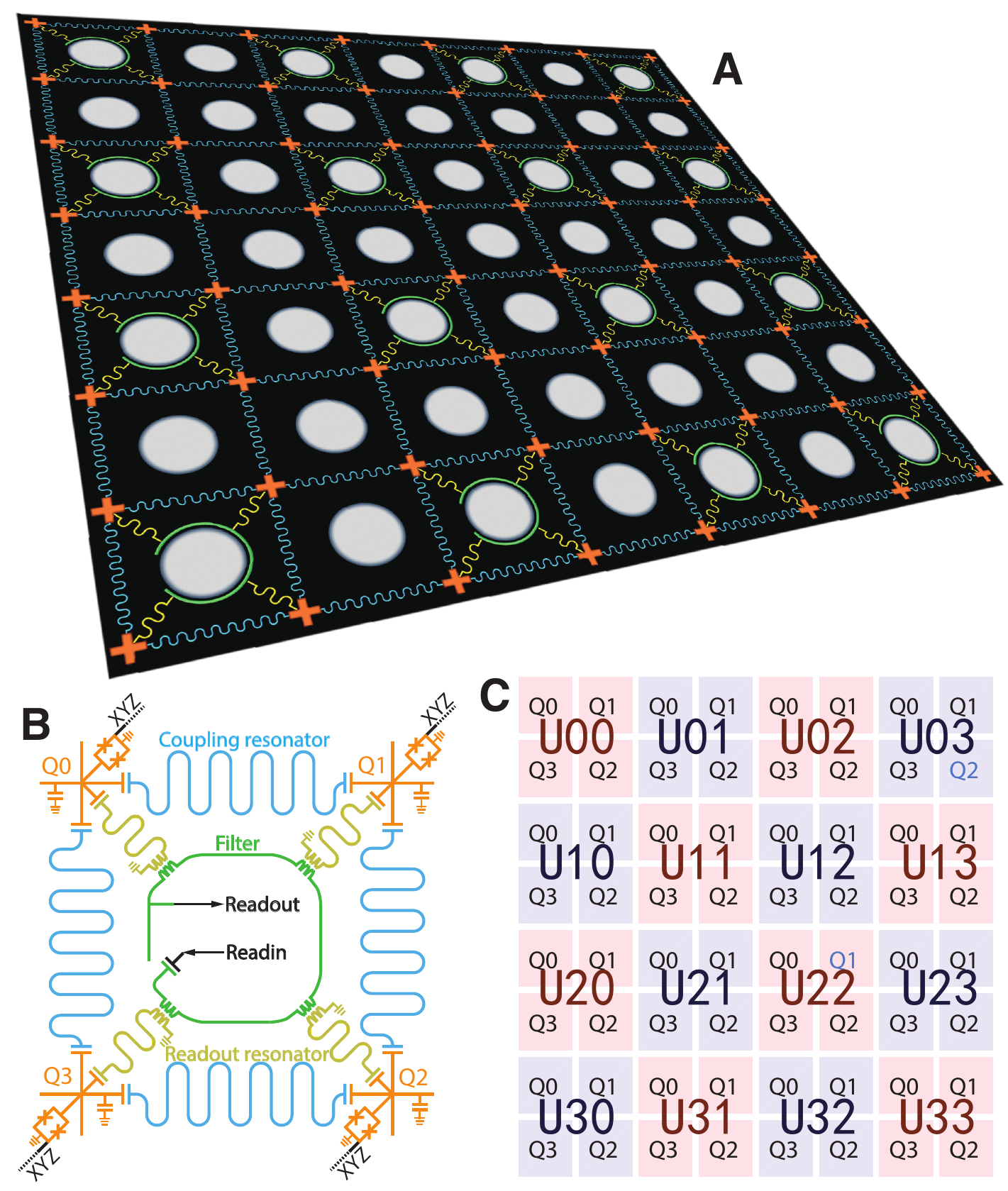}
		\caption{\textbf{The {layout and architecture} of the superconducting quantum {processor}.} (\textbf{A}) {The schematic diagram of the 2D superconducting quantum processor. The orange crosses represent the qubits arranged in an 8$\times$8 array. The gray circles are the pass through holes~\cite{seesm} for 3D wiring. The electrodes for wiring are not shown for simplification.} (\textbf{B}) The circuit diagram of a unit {of} the qubit array.  Each qubit (orange) has an $XYZ$ control line (black) for microwave and pulse control. The qubit couples to an individual $\lambda/4$ readout resonator (yellow) which in turn are commonly coupled to a filter (green). Two neighboring qubits are dispersively coupled to each other through a $\lambda/2$ coupling resonator (blue). (\textbf{C}) The labels of  qubits. Two broken qubits, namely U03Q2 and U22Q1, are marked in blue. 
		}
		\label{fig1}
	\end{figure}

	In our work, we started with the design of a {moderate scale 2D superconducting qubit array.
	One immediately notices the problem associated with planar wiring and how it can be realized to control all the qubits as the size of the 2D array increases.
	One solution has been 3D wiring using techniques like ‘flip chip’~\cite{foxen2017qubit} or ‘through-silicon vias’~\cite{vahidpour2017superconducting}.
	In this work, we provide an alternative technical-friendly solution {based on `pass through holes'}\red{~\cite{seesm}}. This is applied to an 8$\times$8 qubit array (Fig.~\ref{fig1}{A}) that is composed of 16 units whose circuit diagram is shown in  Fig.~\ref{fig1}{B}\red{. Two of the qubits U03Q2 and U22Q1 (Fig.~\ref{fig1}{C}) and one coupling resonator (between U10Q0 and U10Q3) are non-functional.}}
	{We represent the} effective Hamiltonian of the qubit system using the Bose-Hubbard model {as}:
	\begin{eqnarray}
		\widehat{H}&=&\sum_{j\in\{Q_i\}} \hbar \omega_j\widehat{a}_{j}^{\dag }\widehat{a}_{j}+
		\frac{\hbar U_j}{2}\widehat {n}_{j}(\widehat{n}_{j}-1)\cr
		&+&\sum_{{j\in\{Q_i\}, i\in \{C_{Q_i}\}}}
		\hbar  J^{i,j}_{\rm eff}(\widehat{a}_{i}^{\dag}\widehat{a
		}_{j}+\widehat{a}_{i}\widehat {a}_{j}^{\dag}),
		\label{ham1}
	\end{eqnarray}
	where $\widehat{a}_{j}^{\dag }$ ($\widehat{a}_{j}$) are the {qubits} bosonic creation (annihilation) operator, $\omega_j$ the {$j^{\rm th}$} qubit frequency, $U_j$ the anharmonicity, and {$J^{i,j}_{\rm eff}$} the effective coupling strength between $Q_i$ and $Q_j$ \red{via a large detuned coupling resonator,} {where $i$ labels the qubit in group $\{C_{Q_i}\}$ which couples to $Q_i$. }
	For the realization of continuous-time quantum walks (CTQWs), we tune all qubits to the same interaction frequency for time-independent evolution, and the effective evolution Hamiltonian is given by: 
	\begin{equation}
    	\widehat{H}_{\rm evo}=\sum_{i\in\{Q_i\}, j\in \{C_{Q_i}\}}  \hbar J^{i,j}_{\rm eff}(\widehat{a}_{i}^{\dag}\widehat{a
    	}_{j}+\widehat{a}_{i}\widehat {a}_{j}^{\dag})
    	\label{ham2}
	\end{equation}
	{which} forms  an interference network. 
	{By setting the qubits at the} interaction frequency {of} 5.02 GHz, we determined the effective coupling strengths {$J^{i,j}_{\rm eff}$ by} measuring two-qubit swapping oscillations and established ${J_{\rm eff}}/2\pi=2.01\pm0.07$ MHz. 
	
\begin{figure}[H]
		\centering 
		\includegraphics[width=0.9\textwidth]{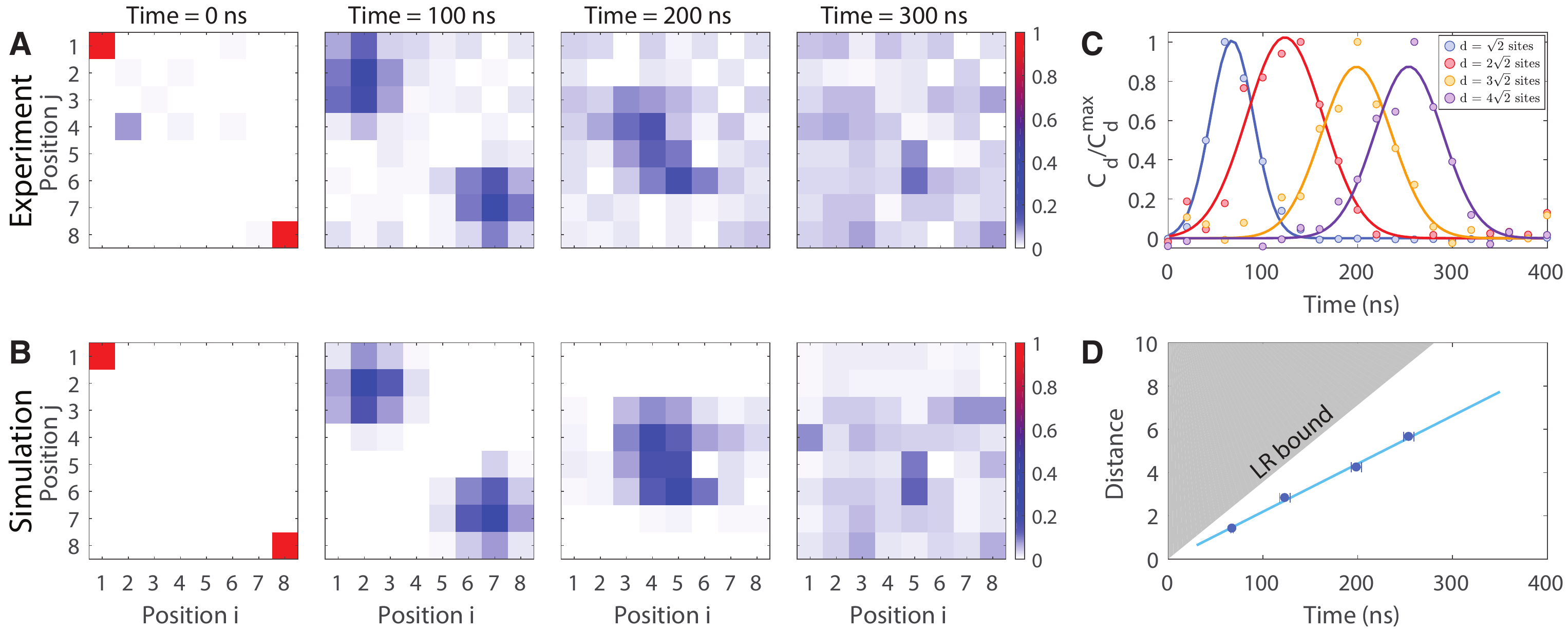}
		\caption{\textbf{Quantum walks on a 2D superconducting qubits array}. 
			(\textbf{A}) The evolution of the measured populations $\langle n_{j} \rangle$ of all qubits at times $t=0$ ns, 100 ns, 200 ns, and 300 ns, respectively, with the two walkers initialized \red{on qubits U00Q0 and U33Q2}. 
			(\textbf{B}) Numerical simulation {of the qubits population evolution under the  same} conditions as (\textbf{A}). 
			(\textbf{C}) The correlation function as a function of time in the case of single-particle QW. The blue, red, orange, and purple circles {represent the measured data points} for the correlation function between the initial excitation site and the sites on the diagonal with distance $d=\sqrt{2}$, $2\sqrt{2}$, $3\sqrt{2}$, $4\sqrt{2}$ sites, respectively. 
			The corresponding solid curves are Gaussian fittings to the data, and the propagation fronts are the center of the Gaussian fittings.
			(\textbf{D}) \red{The propagation velocity and} the Lieb-Robinson bounds. 
			\red{Using a linear fit (blue line) of the propagation fronts (blue circles) with distance, 
			we determine the propagation velocity as $22.2\pm2.0$ site/$\mu s$.}
			The gray shadow shows Lieb-Robinson bound with $v_{max} = 35.7$ site/$\mu s$.
		}
		\label{fig2}
	\end{figure}

	{We begin by exploring continuous-time quantum walks using one and two walkers \red{by exciting one or two qubits on U00Q0 and U33Q2.}
	Once the initial states are prepared, we tune all qubits to the interaction frequency {and allow the system to naturally evolve under Eq.(\ref{ham2}) for a certain time. We then measure the population $\langle \hat{n}_j \rangle$ of all 62 qubits in their $\sigma_z$ basis for} evolution times ranging from 0 to 600 ns. 
	For each time point, we performed 50,000 single-shot measurements. 
	In Fig.~\ref{fig2}{A} we present the experimental results for the two-walker QW with a comparison from numerical simulations (Fig.~\ref{fig2}{B}). \red{In the Supplementary Materials~\cite{seesm} we show the results for the single-walker QWs and the fidelity of the 62-qubit evolution as a function of time.}
	\red{The high fidelity evolutions indicate the high accuracy characterization and high precision control of our system.}
	
	{For quantum walks it is also extremely interesting to determine the  propagation \red{velocity} of the walker (s) through the network compared} to the Lieb-Robinson (LR) bound~\cite{lieb1972finite}. {Focusing on the single walker situation for simplicity, we use the} two-site correlation function defined {by} $C_{ij}(t) = \langle\hat{\sigma}_{z}^{i}\hat{\sigma}_{z}^{j}\rangle-\langle\hat{\sigma}_{z}^{i}\rangle\langle\hat{\sigma}_{z}^{j}\rangle$~\cite{yan2019strongly} {to achieve this (Fig.~\ref{fig2}{C}). 
	{In Fig.~\ref{fig2}{D}, we determine the propagation velocity as} {$22.2\pm2.0$ site/$\mu s$}. 
	{The} maximal group velocity {for two-dimensional systems~\cite{cheneau2012light,takasu2020energy}} is given by {$v_{max}=2\sqrt{2}J_{\rm eff}(1-16J_{\rm eff}^2/9U^2)$} {which equates to $v_{max} = 35.7\;  {\rm site}/\mu s$ in} our system. {Now $v<v_{max}$ clearly shows that our propagation \red{velocity} is limited by the LR bound. \red{The difference is attributed to the short distance and disorders~\cite{cheneau2012light,burrell2009information,seesm}}. }}}

\begin{figure}[H]
		\centering 
		\includegraphics[width=0.9\textwidth]{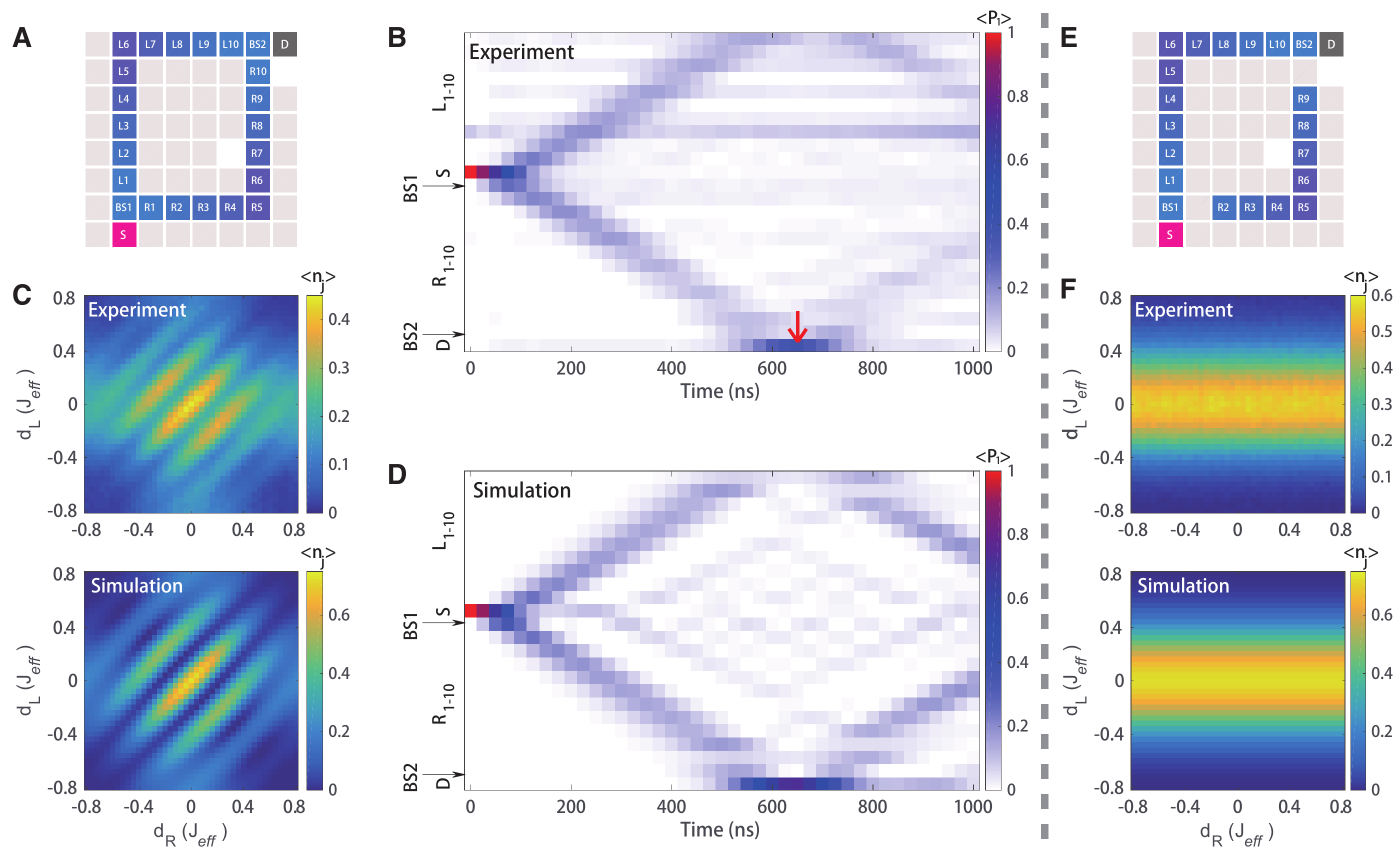}
		\caption{\textbf{Single-particle Mach-Zehnder interferometer}. 
			(\textbf{A}) The {circuit} diagram of the programmable paths for {the realization of} the single-particle Mach-Zehnder interferometer in {a} qubit array. 
			In (\textbf{B}) and (\textbf{D}) we illustrate the dynamics evolution of population $\langle n_{j} \rangle$ of all \red{relevant} sites in experiment and simulation, respectively. The red arrow marks the time $t=650$ ns when the population of $D$ is maximized. (\textbf{C}) The \red{experimental and simulated} population of $D$ at $t=650$ ns under different disorder steps in two paths. (\textbf{E}) The circuit diagram of the interferometer with {the} $\{R\}$ path blocked at $R_1$ and $R_{10}$.
			(\textbf{F}) The \red{experimental and simulated} population of $D$ under different disorder steps in two paths with $\{R\}$ path blocked.
		}
		\label{fig3}
	\end{figure}

	{The CTQWs demonstration establishes a solid basis for the realization of programmable QW. Furthermore, our ability to accurately vary the frequency of each qubit enables us to define propagation paths for the quantum walkers. This is critical for QW-based quantum computing} where we need to deal with graph problems with different structures. 
	In Fig.~\ref{fig3}{A}, we define two intersecting paths {in our 62-qubit superconducting processor} to demonstrate a Mach-Zehnder (MZ) interferometer, where the qubits in the path are tuned to the interaction frequency of 5.02 GHz, while those not involved are biased to 4.97 GHz. 
	After exciting the site $S$, the {walker} will propagate to $BS1$ {where it is split and transmitted} along two {spatially separated paths ($L_1$ to $L_{10}$) and ($R_1$ to $R_{10}$)}. These paths are {reconnected} at $BS2$ {from which the walker arrives at} site $D$.
	The time evolution of all sites' population is measured from $t=0$ to 1000 ns (Fig.~\ref{fig3}{B}). {It clearly shows the single walker traversing both the $\{L\}$ and $\{R\}$ paths.} At $t=650$ ns, a refocusing of the QW with the population as high as 0.43 is observed. An {excellent agreement is found} compared with the numerical simulations (Fig.~\ref{fig3}{D}). 
	
	{Our flexibility in adjusting the qubit frequencies provides another freedom we can exploit associated with} the phase on paths, \red{which is realized with the change of disorders.} {For the $\{R\}$ path elements we adjust the disorder of the sites $R_1$ to $R_5$ from $d_R$ to $5d_R$ respectively, while for sites $R_6$ to $R_{10}$ we did the opposite {changing from $5d_R$ to $d_R$.}} {Similar disorder changes (scaling as $d_L$) are made in the $\{L\}$ path.} By controlling the {disorder sizes}, we measured the population on site $D$ at $t=650$ ns, and observed interference fringes (Fig.~\ref{fig3}{C}). To confirm the origin of {these} fringes, {we blocked the path of $\{R\}$ on $R_1$ and $R_{10}$ (Fig.~\ref{fig3}{E}) and found no interference fringes (Fig.~\ref{fig3}{F}).} Such results show that the disorder not only {changes} the tunneling amplitude between neighboring sites, but also provides the quantum walker a different phase accumulated in propagation {that gives rise to} the interference fringes. {Moreover, for that interference to have occurred, our walker must have maintained coherence as it traverses a superposition of distinct spatially separated paths.}
	{The generation of those non-local correlations is essential for the development of QW based universal quantum computation.} 

\begin{figure}[H]
		\centering 
		\includegraphics[width=0.9\textwidth]{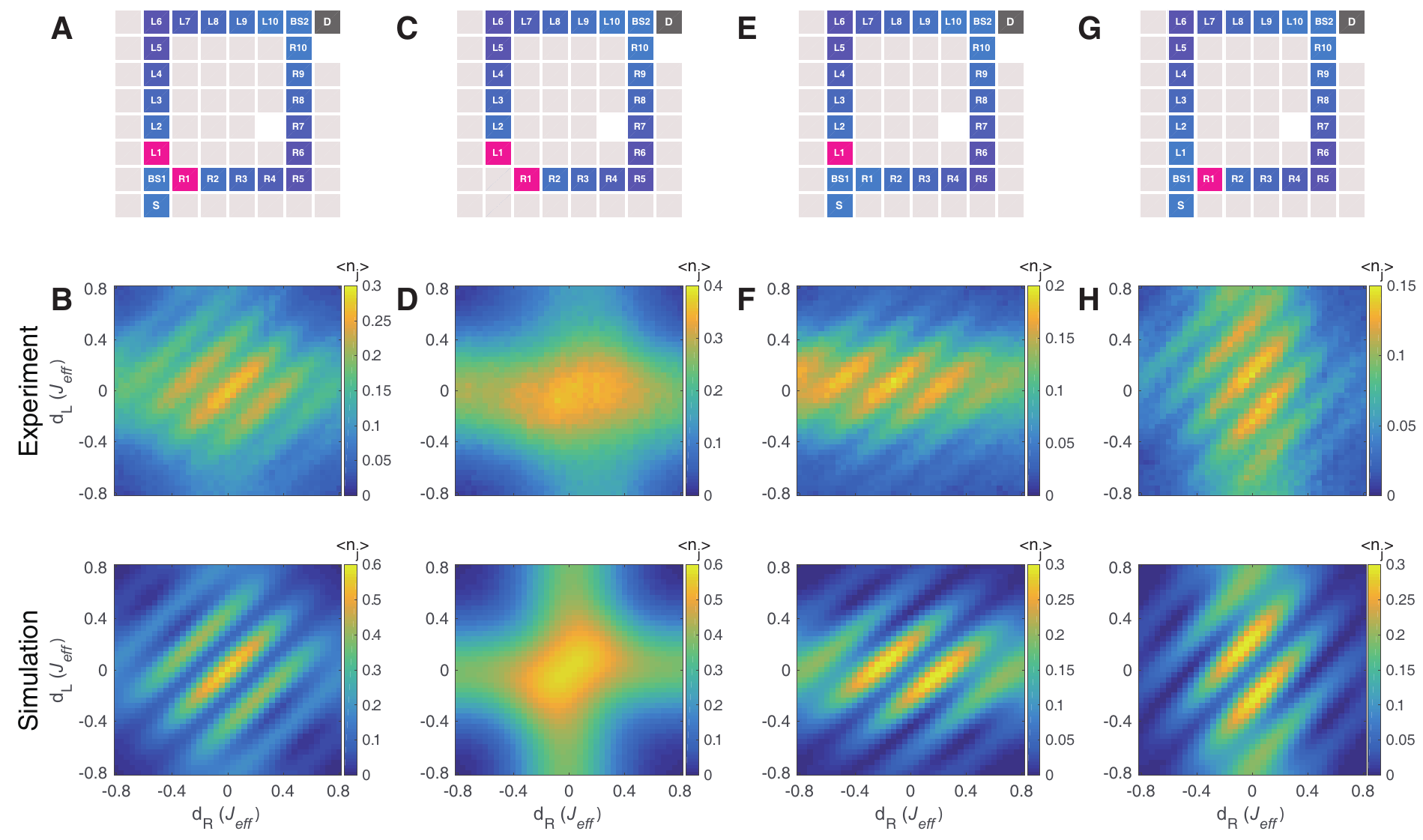}
		\caption{\textbf{Two {walkers} in the Mach-Zehnder interferometer.} 
			(\textbf{A, C, E, G}) Circuit diagrams of the programmable paths for the Mach-Zehnder interferometer. The initial state {composed of either a single or double walker} is prepared by the excitation of the sites marked with pink. 
			(\textbf{B,D,F,H}) The \red{experimental and numerically simulated} population of $D$ at $t=550$ ns are shown for the various \red{configurations below the circuit diagram. } 
		}
		\label{fig4}
	\end{figure}

	{The natural question that now arises is what occurs when we have multiple walkers in our MZ interferometer. 
	We then create two walkers on sites $L_1$ and $R_1$ (Fig.~\ref{fig4}{A}) by exciting these respective qubits, and then let the system evolve. We} measured the population on site $D$ after $t=550$ ns, and {observed the interference fringes} in Fig.~\ref{fig4}{B}. {This is a similar pattern to what we observed in the single walker case. To determine the origin of this interference fringes in the two walker case, we performed a number of control experiments beginning with the removal of sites} $BS1$ and $S$ (Fig.~\ref{fig4}{C}), which stops the both walkers back propagating to take their alternate path. As shown in Fig.~\ref{fig4}{D}, no interference fringes {are} observed anymore. 
	This indicates that the pattern comes from the interference between the single-particle forward and back propagation. {Next, we created a single walker at either site $L_1$ or $R_1$ (Figs.~\ref{fig4}{E}/{G}) respectively, and let it walk through the interferometer.}
	The results both clearly show interference fringes (Figs.~\ref{fig4}{F}/{H}). However, neither {one of them nor their sum~\cite{seesm} are} the same as {what we observed in} Fig.~\ref{fig4}{B}. 
	{This re-enforces our observation that the two walkers present in the MZ interferometer must have interacted with each other.}
	Such results agree well with our understanding of transmon qubit physics in the hard-core boson limit~\cite{lahini2012quantum}, where $|U/J_{eff}|\sim120$.

	{To summarize},	our successful demonstration of quantum walks in two dimensions and the corresponding realization of MZ interferometers clearly shows the potential of these superconducting qubit processors. With the remarkable control of not only the qubits frequencies but also the tunneling amplitude and phase between neighboring sites, these superconducting-circuit based quantum walks are  elegant approach for the exploration of hard-core boson interference beyond that achievable in photonic systems. Furthermore, multi walker realizations will push us into the quantum advantage realm as {the excitation number and/or the} processor size increases. Finally, the demonstration of programmable quantum walks {in superconducting quantum processors} is an essential {technological} milestone {providing a} solid basis for more complex quantum many-body simulations, and {which in the future} can be further applied to quantum search algorithms and even universal quantum computing.

	\bibliographystyle{Science}

	\clearpage
	
	\section*{Acknowledgments}
	The authors thank the USTC Center for Micro- and Nanoscale Research and Fabrication for supporting the sample fabrication. 
	The authors also thank QuantumCTek Co., Ltd., for supporting the fabrication and the maintenance of room-temperature electronics.
	\textbf{Funding:}
	This research was supported by the National Key R\&D Program of China (Grant No. 2017YFA0304300), the Chinese Academy of Sciences, Anhui Initiative in Quantum Information Technologies, Technology Committee of Shanghai Municipality, National Science Foundation of China (Grants No. 11905217, No. 11774326), Shanghai Municipal Science and Technology Major Project (Grant No. 2019SHZDZX01), Natural Science Foundation of Shanghai (Grant No. 19ZR1462700), and Key-Area Research and Development Program of Guangdong Provice (Grant No. 2020B0303030001). This work was also supported in part by the Japanese MEXT Quantum Leap Flagship Program (MEXT Q-LEAP), Grant No. JPMXS0118069605. 
	\textbf{Author contributions:}	
	X.Z. and J.-W.P. conceived the research. M.G., S.W, C.Z., M.-C.C., and X.Z. designed the experiment. S.W., M.G., Q.Z., Y.Z., Y.Y., F.C., C.Y., and X.Z. designed the sample. S.W., C.Z., H.R., H.D., K.Z., S.C., and Y.-H.H. prepared the sample. S.G., H.Q., and H.D. prepared the Josephson parametric amplifiers. Y.W. developed the programming platform for the experiments. M.G., S.W., C.Z., Y.Z., S.L., C.Y., J.Y., D.F., D.W., and H.S. contributed to the building of the ultra low-temperature and low-noise measurement system. J.L., Y.X., F.L., C.G., L.S., N.L., and C.-Z.P. developed the room-temperature electronics. M.G., S.W., C.Z., M.-C.C., H.-L.H., V.M.B., K.N., W.M., C.-Y.L., and X.Z. performed the data analysis. All authors contributed to the discussions of the results and the preparation of the manuscripts. X.Z. and J.-W.P. supervised the whole project.
	\textbf{Competing interests:} None declared.
	\textbf{Data and materials availability:} All data needed to evaluate the conclusions in the paper are present in the paper or the supplementary materials.
	
	\section*{Supplementary materials}
	
	 \noindent Materials and Methods \\
	 Supplementary Text \\
	 Figs. S1 to S13\\
	 Table S1 \\
	 Movie S1 to S4\\
	 References\\
	
\end{document}


\clearpage
	
	\maketitle
	
	\clearpage
	
	
	\tableofcontents
	
	\section{Experimental wiring setup}

\begin{figure*}[htb!]
		\centering \includegraphics[width=0.7\linewidth]{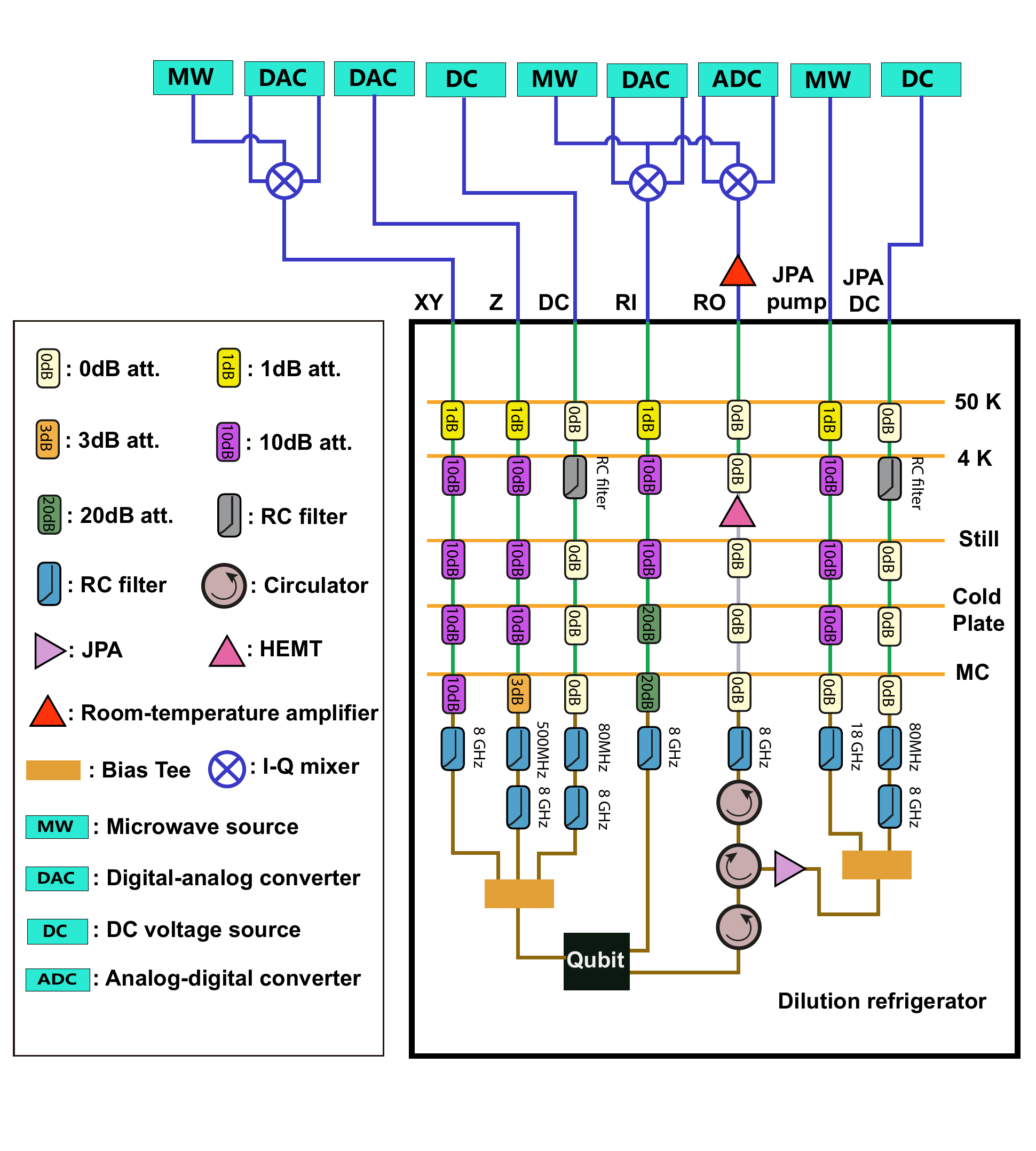}
		\caption{The schematic diagram of control electronics and wiring. For each qubit, there are individual $XY$, $Z$ and DC control lines, which are combined together via bias tees before connected to the quantum processor. In the dilution refrigerator, attenuators and filters are installed at various stages to reduce noise. The Josephson parametric amplifiers (JPAs) and high electron mobility transistors (HEMTs) are used to amplify the readout signals. At room temperature, digital to analog converters (DAC) and microwave sources are used to generate pulses for qubit $XY$ control and readout. Qubit $Z$ control pulses are generated by DACs. DC voltage sources are used to bias the qubits to their idle points and to bias the JPAs. The readout signals amplified by the room-temperature amplifiers can be digitized and demodulated by analog to digital converters (ADCs).}
		\label{figSa}
	\end{figure*}

	As shown in Fig.\ref{figSa}, the 62-qubit processor is installed at the base temperature stage of the dilution refrigerator (DR), which is cooled down to 10 mK. In the DR, we used  totally 186 control lines for qubits, 32 control lines for Josephson parametric amplifiers (JPAs), 16 readout input lines and 16 readout output lines. For each qubit, there are three control lines, the $XY$ line for qubit driving, the fast $Z$ control line to apply the $Z$ pulse control, and the DC line to bias the qubit to its idle point. The four qubits in a unit share one readout input line and one readout output line.
	
	In the DR, attenuators and filters are installed at different stages to reduce noise. To reduce the thermal noise from higher-temperature stages, there are totally 41 dB, 34 dB and 61 dB attenuations for $XY$ control line, fast $Z$ control line and readout input line, respectively. In addition, at the base temperature stage, we installed 8 GHz low pass filters for all $XY$, fast $Z$, DC control lines, readout input and readout output lines, 500 MHz low pass filters for fast $Z$ control lines and 80MHz low pass filters for DC control lines, to further reduce the high-frequency noise. At the 4 K stage, we installed a RC filter of 10 KHz cut-off frequency for each DC control line. After passing through the attenuators and filters, the $XY$, fast $Z$ and DC signals are combined together by a bias tee at base temperature and then reach the quantum processor. For qubit state readout, the readout input signal passes through the attenuators and the 8 GHz low pass filter and then reaches the quantum processor. The output signal firstly passes through two circulators, then amplified by a Josephson parametric amplifier (JPA), of which the noise as well as the noise from higher temperature stages has been blocked by the preceding of two circulators. Then the signal passes through a third circulator and an 8 GHz low pass filter, amplified by a high electron mobility transistor (HEMT) amplifier at the 4K stage and a low noise amplifier at room temperature respectively, and finally captured and analyzed by the room temperature electronics. To reduce the noise in controlling JPA, we installed 31dB attenuators and a 18 GHz low pass filter for each JPA pump line, a RC filter, an 8 GHz and an 80 MHz low pass filter for each JPA bias line.
	
	At room temperature, we use two digital to analog converters (DAC) channels to generate Gaussian shaped pulses for each $XY$ control. These two intermediate frequency pulses are up-converted to the driving frequency by an IQ mixer with a carrier frequency of 5.72 GHz generated by the microwave source. We use one DAC channel for the $Z$ pulse control and one DC source for the DC bias. For the readout of qubit state, we use two DAC channels and one microwave source to generate a multi-tone readout input signal by side-band mixing with an IQ mixer. The readout output signal is down-converted into two signals, which are digitized and demodulated by the two ADC channels to extract the qubit state information.

	\section{{Device design and fabrication}}
	\red{In the main text, the quantum processor illustrated in Fig. 1A contains an 8$\times$8 qubit array which is composed of 16 units whose circuit diagram is shown in  Fig. 1{B}. Here each unit contains four frequency tunable transmon qubits which are dispersively coupled to individual readout resonators but share a common band-pass filter for state readout. Each qubit couples to its four nearest neighbors via coplaner waveguide resonators. The separation between qubits is approximately 4 mm, leaving sufficient space to cut {holes in the chip substrate} using a picosecond laser. The control lines and readout filters are connected to the fan-out PCB on the bottom of the chip by wire bonding through the holes, {from which they} can be further connected to the coaxial cables installed in the dilution refrigerator.  Parasitic slot-line modes, which may arise,  are then suppressed using air-bridges applied over the control lines, readout resonators, filters and coupling resonators to connect the ground planes.
	
	The quantum processor is fabricated with the following steps:
    \begin{enumerate}
    \item{ First a 100 nm Al film is growed on a 2 inch sapphire wafer using molecular beam epitaxy.}
    \item{ Optical lithography and wet etch are then applied to define the resonators, filters, control lines and trasmon capacitors.}
    \item{ Optical lithography and deposition of Ti and Au films are then used to define the alignment marks for electron-beam lithography.}
    \item{ Next the dielectric scaffolds for airbridges are defined by optical lithography and deposition of SiO$_{2}$ film.}
    \item{ The Al airbridges are defined by optical lithography and deposition of Al film.}
    \item{ Electron-beam lithography, followed by double-angle evaporations of Al, is applied to fabricate the qubits’ Al/AlO$_{x}$/Al Josephson junctions.}
    \item{ The 2 inch wafer is diced into a square chip and pass through holes are fabricated by laser.}
    \item{ A VHF etcher is used to release the airbridges by removing the scaffolding SiO$_{2}$.}
    \item{ The quantum chip is connected to the PCB by wire bonding.}
    \end{enumerate}
	}

	\section{Parameters of the superconducting quantum device}
	
    \begin{table*}[htb!]
    \centering
    \begin{tabular}{cccc}
    \toprule
    Parameters& Median& Mean& Stdev.\\
    \midrule
    Qubit maximum frequency (GHz)& 5.434& 5.442& 0.116\\
    Qubit idle frequency (GHz)& 5.148& 5.200& 0.198\\
    Qubit anharmonicity $\eta/2\pi$ (MHz)& -251.0& -248.9& 7.0\\
    $T_1$ at idle frequency ($\mu$s)& 12.48& 13.56& 5.53\\
    $T_1$ at working point ($\mu$s)& 11.24& 12.26& 5.45\\
    $T_2^*$ at idle frequency ($\mu$s)& 1.61& 1.63& 0.67\\
    Coupling strength between qubit and readout resonator (MHz)& 95.49& 94.95& 4.60\\
    Effective coupling strength between neighboring qubits (MHz)& 1.99& 2.01& 0.07\\
    Dispersive shift $\chi/2\pi$ (MHz)& 1.05& 1.14& 0.34\\
    Resonator linewidth $\kappa/2\pi$ (MHz)& 4.91& 5.06& 1.63\\
    Readout fidelity of $\ket{0}$& 0.970& 0.966& 0.016 \\
    Readout fidelity of $\ket{1}$& 0.932& 0.919& 0.032 \\
    Effective qubit temperature (mK)& 65& 66& 11\\
    \bottomrule
    \end{tabular}
    \caption{Statistics of the qubits' parameters.} 
    \label{tableSa}
    \end{table*}
    
    \begin{figure*}[htb!]
		\centering \includegraphics[width=1\linewidth]{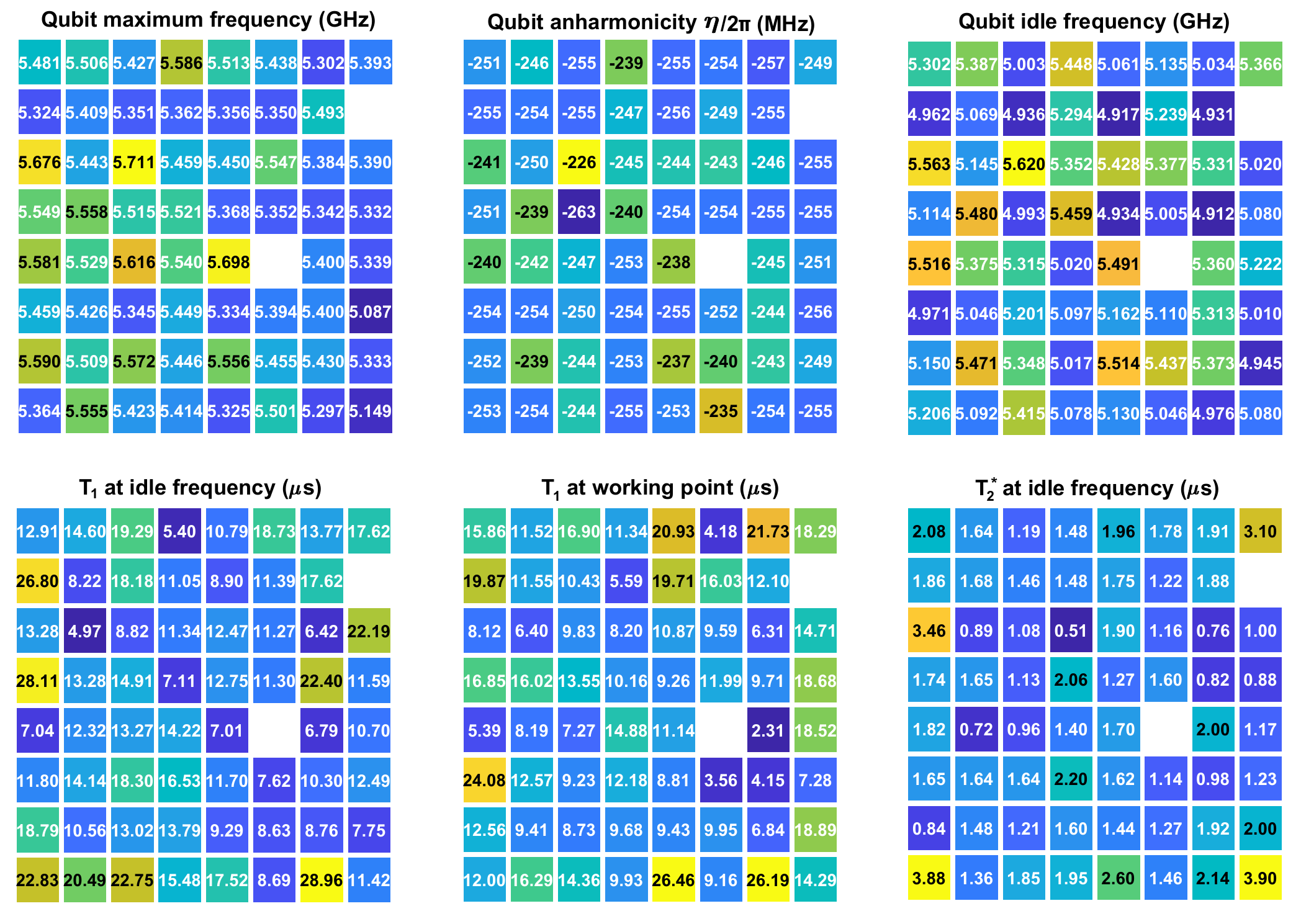}
		\caption{Qubit parameters distributions, including qubit maximum frequency, qubit anharmonicity, qubit idle frequency, qubit energy relaxation time $T_1$ at the idle and working points, and qubit dephasing time $T_2^*$ at idle frequency. Each square in the diagrams represents a qubit, the number and color in the square show the value of the corresponding parameter.}
		\label{figSb}
	\end{figure*}
    
    \begin{figure*}[htb!]
		\centering \includegraphics[width=1.0\linewidth]{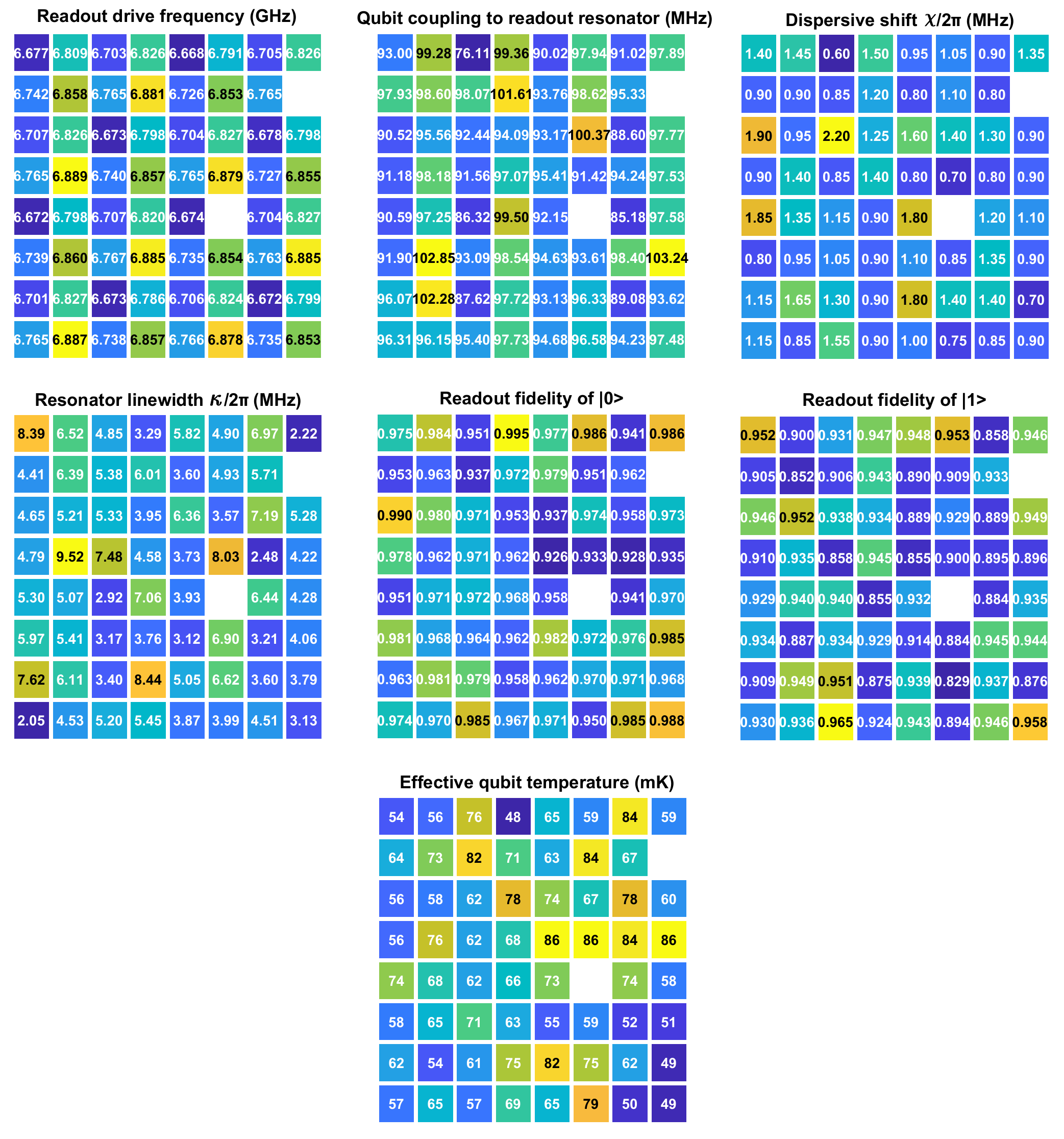}
		\caption{Qubit readout parameters distributions, including readout drive frequency, the coupling strength between the qubit and its individual readout resonator, the dispersive shift, the resonator linewidth, the readout fidelity of state $\ket{0}$, the readout fidelity of state $\ket{1}$ and the effective qubit temperature. }
		\label{figSc}
	\end{figure*}

     \begin{figure}[htb!]
		\centering \includegraphics[width=0.5\linewidth]{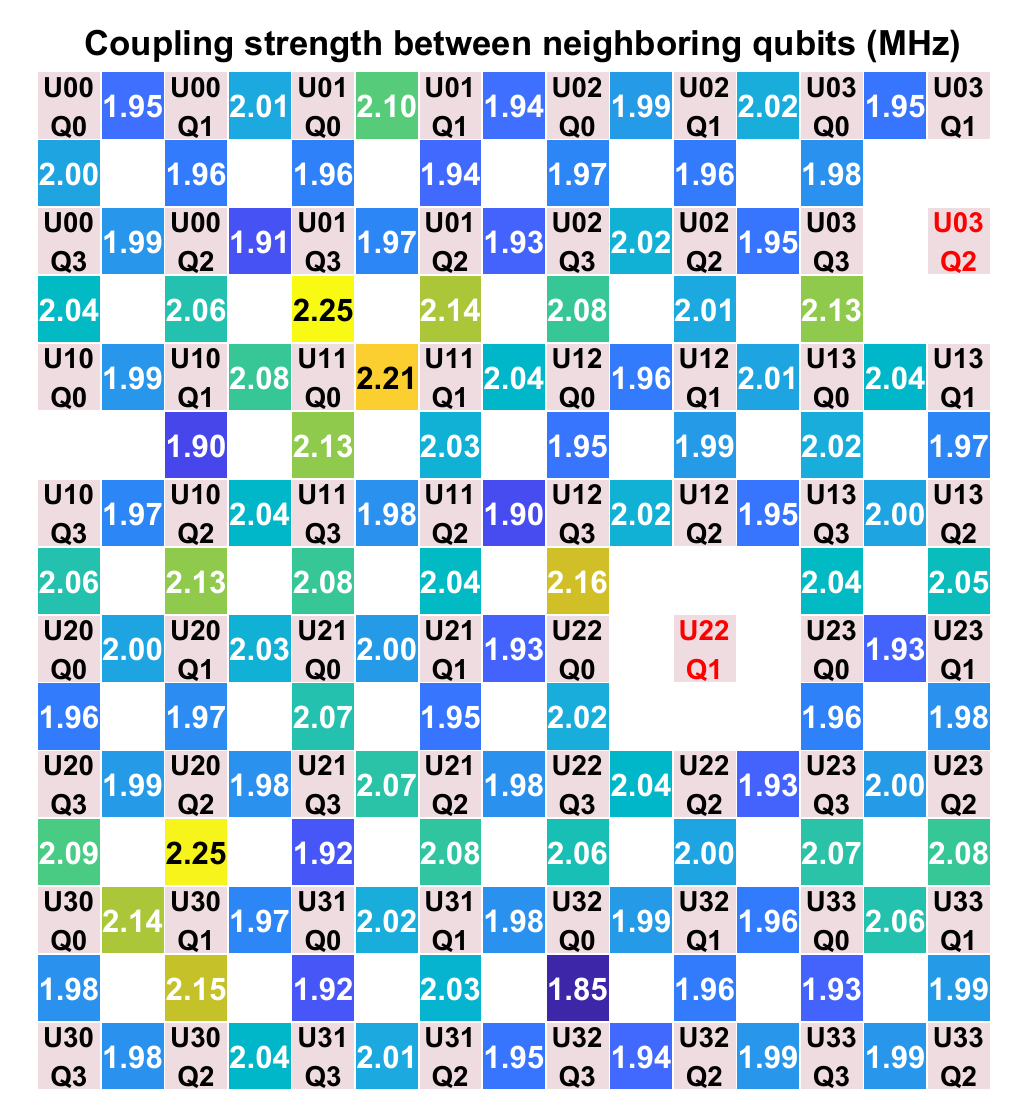}
		\caption{Distribution of the coupling strengths between neighboring qubits. The square connecting two qubits shows the effective coupling strength between them when these two qubits are tuned to the interaction frequency of 5.02 GHz.}
		\label{figSk}
	\end{figure}
	
	Our experiments are implemented on a 62-qubit quantum processor arranged in an 8$\times$8 qubits array composed of 16 units, as illustrated in Fig.1 of the main text. The qubits in the processor are frequency tunable transmons, of which the frequencies can be tuned individually. The qubit maximum frequencies range from 5.087 GHz to 5.711 GHz with the standard deviation of 116 MHz.The qubits are biased to the idle points ranging from 4.912 GHz to 5.620 GHz by DC signals. The idle points are chosen to avoid cross-talk between neighboring qubits, reduce the influence of defects on qubits and achieve relatively high qubit energy relaxation time $T_1$. The qubits can be tuned to their working points by $Z$ pulses to realize interaction. The anharmonicity $\eta/2\pi$ of qubit, which is equal to the nonlinear on-site interaction $U/2\pi$ in the Bose-Hubbard Hamiltonian (1) in the main text, has the mean value of -248.9 MHz. Each qubit couples to its four nearest neighbors via coplaner waveguide resonators, whose frequencies are designed to be 6 GHz, and the coupling strength between the qubit and coupling resonator is designed to be around 44 MHz.  The mean value of the measured effective coupling strength between neighboring qubits is 2.01 MHz at the interaction point of 5.02 GHz, which gives $|U/J|=124$. This means our system can be described by hard-core bosons~\cite{lahini2012quantum}. For qubit state readout, the mean value of dispersive shift is 1.14 MHz, and the mean value of  readout resonator linewidth is 5.06 MHz. With these parameters, we achieve an average readout fidelity of 96.6$\%$ for state $\ket{0}$ and 91.9$\%$ for state $\ket{1}$. The effective qubit temperature extracted from the excited state probability in measuring $\ket{0}$ state is determined as 66 mK in average. More details of the parameters for each functional qubit are shown in the Fig.\ref{figSb}, Fig.\ref{figSc} and Fig.\ref{figSk}, and their statistical values are shown in table \ref{tableSa}.
	\clearpage
	\section{System calibration}
	\red{In our experimental setup, the qubits are biased at their idle frequencies for state preparation and readout. We optimized the qubit frequency setup to maximize the energy relaxation time $T_1$ of all qubits while preventing the effect of {defects, $ZZ$ coupling, and} microwave cross-talk. For the realization of continuous-time quantum walks, we tune all qubits to the same interaction frequency for time-independent evolution. To achieve this in practice and ensure its stability, three experimental requirements {need to be realized: (i)} the stability of the qubit frequency, {(ii)} the {precise} knowledge of the coupling strength, and {(iii)} high precision control {of the} qubit-frequency alignment. {This is achieved through a series of calibration experiments. We begin with} a $Z$-pulse distortion calibration \cite{yan2019strongly} to ensure the stability of qubit frequency  {while} performing detuning operations. 
	{Then, by setting the qubits at the} interaction frequency {of} 5.02 GHz, we determined the effective coupling strengths {$J^{i,j}_{\rm eff}$ by} measuring two-qubit swapping oscillations and established ${J_{\rm eff}}/2\pi=2.01\pm0.07$ MHz. 
	{The final requirement is achieved through several rounds of qubit-frequency alignment} calibrations and corrections {which allows us establish that the} disorders of all qubits are no larger than 1.6 MHz ($0.8J_{\rm eff}/2\pi)$. More details of the calibrations are as follows.}
	
	\subsection{Idle frequency setup}
    In calibrating and optimizing qubit idle frequency, there are several key elements we need to focus on. They are:
    
    \begin{enumerate}
    \item{\textbf{Energy relaxation.} It is well known that the energy relaxation time of the qubit varies strongly with frequency. Defects including two-level-systems (TLSs), slot-line modes, and some other microwave modes, induce significantly short energy relaxation times at certain frequencies. This will further limit the readout and single-qubit operation fidelities. Those frequencies affected by defects should be avoided.}
    
    \item{\textbf{Microwave crosstalk.} In our superconducting quantum device, though the microwave crosstalk has been suppressed to below -40 dB for next-nearest-neighboring qubits, for the nearest-neighboring qubits it is still not negligible as being around -25 dB. Considering the maximal driving strength to be around 25 MHz, the minimal frequency gap between two neighboring qubits is set to be 50 MHz. Meanwhile, to avoid the two-photon excitation by crosstalk, the frequency of exact match with $f_{02}/2$ of the neighboring qubits should be avoided, where $f_{02}$ is the frequency difference between the ground state and the second excited state.}
    
    \item{\textbf{$ZZ$ coupling.} For the coupled qubits, the $ZZ$ coupling strength is given by $\Omega_{ZZ} = -2g^2(\eta_1+\eta_2)/{[(\Delta-\eta_1)(\Delta+\eta_2)]}$, where $\eta_1$ and $\eta_2$ are the qubit anharmonicities, $g$ is the coupling strength between these two qubits, and $\Delta$ is the difference in qubit frequencies. From the above expression, it is noticed that if one wants to limit the $ZZ$ coupling to be below 0.2 MHz, the frequency gap between $f_{01}$ of the target qubit and $f_{12}$ of the coupling qubit should be larger than 45 MHz, where $f_{12}$ is the frequency difference between the first and second excited states. }
    
    \item{\textbf{Readout fidelity.} We perform the readout operation at the idle frequency. The large detuning from the readout resonator reduces the dispersive coupling strength of the qubit, and further affects the readout fidelity. However, we also need to balance it with the frequency crowding issues. We set the minimal idle frequency to be 4.9 GHz.}
    
	\end{enumerate}
	
    We use this principle to construct a three step optimization procedure: 
    
    \begin{enumerate}
    \item First, we measure the frequency-dependent $T_1$ of all qubits to determine the defect-affected frequencies. We then generate tables of available frequencies $\{f_{avl}^{i}\}$ for all qubits by removing those bad-performance points. The frequency step in the tables is 1 MHz.
    
    \item Second, based on $\{f_{avl}^{i}\}$, we search for a solution that all qubits are initialized in their available frequencies. To be more specific, we randomly choose one frequency from $f_{avl}^{q}$ as the initial frequency for the qubit $q$ with the shortest length of $f_{avl}^{q}$. Then, we update $\{f_{avl}^{i}\}$ with the frequencies of determined qubits, following the principles we listed previously. We repeat this progress until all qubits are set. However, if there is no available frequencies for one of the qubits, we restart the search again. We note that we use $T_1$ as the weight of each frequency point, thus those frequencies with better performance have more chances to be chosen. 
    
    \item Finally, we optimize all single-qubit performance in parallel, and then measure $T_1$ of all qubits at their idle frequencies. It is still possible that when the qubits are biased to different idle points, some microwave modes and TLSs may be changed, resulting in different available frequencies. The changes can be identified from the relatively low values and the large variations in fitting $T_1$. We then optimize those qubits again by measuring $T_1$ at different available frequencies to find a frequency with better performance. This step is repeated for several rounds until the performance of all qubits are acceptable. 
    \end{enumerate}
    
	\subsection{Frequency-alignment optimization}
	In our previous experiments~\cite{yan2019strongly}, we optimized the frequency-alignment by tuning all qubits to the same interaction frequency for evolution, and then used the population propagation to fit the disorders on each site for further corrections. However, when the system size grows to be more than 60 qubits, such a strategy becomes unachievable because of the non-negligible time cost in the numerical simulations. 
	
	In this work, we optimize the frequency-alignment with the following procedure:
	
	\begin{enumerate}
	    \item{Set the initial alignment frequency for all qubits. In this work, it is 5.02 GHz. }
	    
	    \item{For each qubit, we measure the population propagation between the qubit and its coupled qubits, which is defined as a `multi-qubit swapping' experiment. We begin by exciting a single qubit. Then we tune the qubit and its coupled qubits to the alignment frequency for system evolution. All other qubits are tuned to 4.97 GHz to prevent any unwanted state leakage. After an evolution time ranging from 0 to 1 $\mu s$, we tune these qubits back to their idle points to readout the population of each qubit jointly. }
	    
	    \item{With the data sets for all qubits, we use the `Nelder-Mead' algorithm to search for a disorder map which has the best fit to the data. To be more specific, for each data set, we can calculate the distance from the experimental data to the numerically simulated data with the disorders given by the disorder map. By defining the cost function as the sum of the square of the distances, we optimize the disorder map to minimize the cost function. The disorder map is the final frequency differences we need to correct the alignment frequency for each qubit. Meanwhile, we define and calculate the overall distance using the sum of the square of the distances, which comes from the data and the numerical simulations with no disorders.}
	    
	    \item{We then determine the sign of correction by adding or subtracting the disorder map for the alignment frequency. Then, there are two alignment frequency setups, of which one setup is worse and will induce larger overall distance. By running `multi-qubit swapping' experiments for both setups, we compare the overall distances. The smaller overall distance indicates the better setup, and can be used for further updating. }
	    
	    \item{By repeating steps 2 to 4 for several rounds, the overall distance will saturate. In our case the final maximal disorder determined is smaller than 0.8J$_{eff}$.}
	    
	\end{enumerate}
	
	\subsection{Optimization of the interferometer}
	Though the disorders of each site have been suppressed to be below 0.8J$_{eff}$, the residual disorders in the interferometer are still non-negligible. As the number of effective coupling is reduced from 4 to 2 or 3 in the interferometer, the disorder causes the reflection in spreading and reduces the effective coupling strength. As a result, the maximal population at site $D$ at $t=650$ ns is only 0.12 before optimization.
	
	The ultimate objective of the optimization is to enhance the population of site $D$ at $t=650$ ns when single excitation is involved in site $S$. However, the direct optimization of all sites in the interferometer may result in a different local minimum, which corresponds to the blockade of one path. Therefore, we use a procedure utilizing a two-step optimization to search for the best correction.
	
	\begin{enumerate}
	    \item First, we optimize the alignment frequency of the interferometer except for sites $BS2$ and $D$. After exciting the qubit at site $S$, we tune all qubits in the interferometer except for sites $BS2$ and $D$ to the alignment frequency. Note that in this step, sites $BS2$ and $D$ are not in interaction. We use the product of the populations of sites $L_{10}$ and $R_{10}$ at $t=550$ ns as the cost function, and use `Nelder-Mead' algorithm to optimize the alignment frequencies of the corresponding qubits.
	    
	    \item Second, we optimize the alignment frequencies for all sites. After exciting the qubit at site $S$, we tune all sites in interaction and then use the population of site $D$ at $t=650$ ns as the cost function. Again, we use the `Nelder-Mead' algorithm to optimize the alignment frequencies for all sites. Now based on the first step optimization, the two paths have been in balance, thus the local minimum with one path blocked will not occur. 
	\end{enumerate}
	
	Following this two-step optimizations, the maximal population at site $D$ for $t=650$ ns can be optimized to be above 0.43. 
	
	\section{Thermalization and Post-selection}
	
	{We note that for several qubits, thermal noise is non-negligible when performing detuning operations, which is possibly caused by the heating of the bonding wires to the electrodes of the control lines. These affects the quality of our quantum walks. }
	
	
	
	\begin{figure*}[htb!]
		\centering \includegraphics[width=0.95\linewidth]{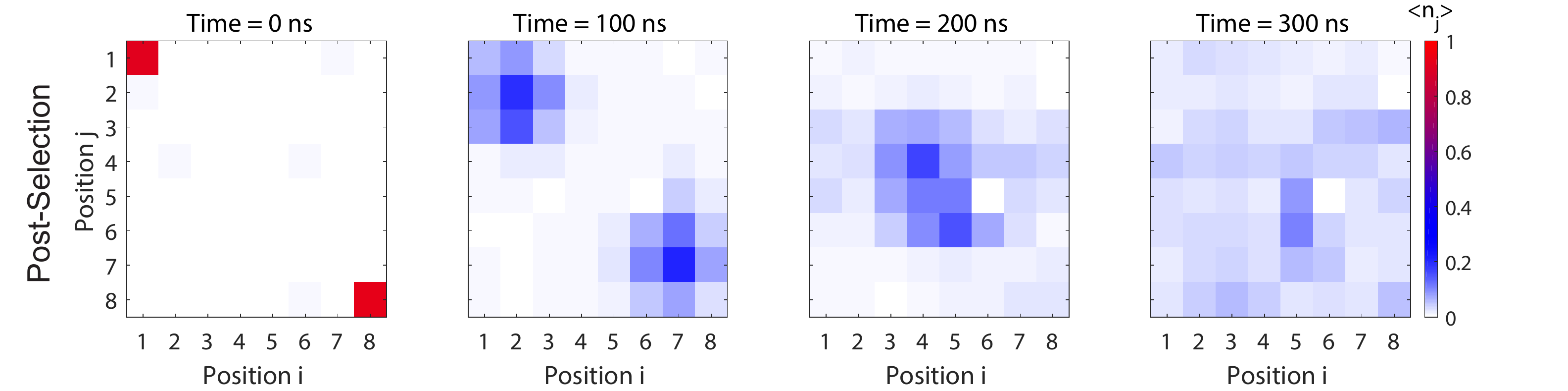}
		\caption{{\textbf{The two-particle quantum walks after post-selection.} For the same set of raw data in Fig. 2{A} in the main text, we performed post-selection with the conservation of the total excitation number. After post-selection, the behavior is much closer to ideal as the thermalization is suppressed. \red{We note that no post-selection is used to any other results in the main text and the Supplementary Materials except for this figure.}}
		}
		\label{figps}
	\end{figure*}
	
	{To suppress the thermal excitation, one solution is to utilize post-selection, a technique commonly used in linear optical quantum computation~\cite{rohde2012error}. Post-selection involves selecting data events where the total number of recorded excitations are the same as the total number of initial excitations. This technique allows us to partially remove the effects of circuit loss, thermalization and detector inefficiency. All of these individually change the total excitation number and so if there is only one error we can eliminate it. However we do not remove the cases where two errors occurs which maintain the excitation number. Examples of these two error processes for instance include when an excitation is loss in the circuit while one detector reports $|1\rangle$ when it was actually $|0\rangle$. Another is when one detector reports $|1\rangle$ (but was $|0\rangle$) while the second reports $|0\rangle$ (but was $|1\rangle$). Both these examples maintain the total excitation number and so can not be distinguished from the ideal case. However two error events occur with a much lower probability than single error events. The improved quantum walks after post-selection with two walkers are shown in Fig.~\ref{figps}. \red{After the post-selection, about 11.3$\%$ of the total number of all single-shot measurements are retained.} We note that for the other results, no post-selection is used.}

	
	\section{Numerical simulation method}
	To obtain the theoretical results \red{of the hard-core Bose-Hubbard model}, as the comparison of the experiment results, we numerically simulate the evolution of the system with 62 functional qubits under the spin model. This is a reasonable approximation of the Bose-Hubbard model when $\vert U\vert \gg \vert J\vert$ along with low filling factor. However, the full Hamiltonian space of the 62-qubit system consists of $2^{62}$ dimensions, whose corresponding matrix has $2^{62}$ bases. The space has been beyond a state-of-the-art classical supercomputer. So we truncate the full Hamiltonian space to a subspace whose bases have the same number of excitations as the initial states. For example, if we have a single excitation in the qubit array, we truncate the full space to the subspace composed of $\left\lbrace \vert 10\cdots0 \rangle,\ \vert 01\cdots0 \rangle,\ \cdots\cdots,\ \vert 00\cdots1 \rangle\right\rbrace$. In this way, we reduce the number of dimensions from $2^{62}$ to $1891$ for two excitations and $62$ for a single excitation. The corresponding matrix of the truncated Hamiltonian is small enough to be easily handled with a laptop. \red{However, in this simplification, it is difficult for us to take decoherence into consideration, as the Hilbert space is not complete any more. Meanwhile, the device is also assumed as an ideal system, with no disorders on qubits' working frequencies, and the coupling strength between neighboring qubits are all assumed as the average of the experimentally measured value. Our Hamiltonian in the simulation represents the ideal Hamiltonian of our experiments without any imperfects.}
	
	\red{With the ideal time-independent Hamiltonian at hand,} we can transform the Schrodinger equation to $\vert \Psi(t)\rangle = e^{-i H t/\hbar}\vert \Psi(0)\rangle$, where $\vert \Psi(0)\rangle$ is our initial state while $\vert \Psi(t)\rangle$ is the state of system at time $t$. In this way, with the matrix of truncated Hamiltonian and the vector of the truncated initial state, we can numerically simulate the evolution of the initial state at any time under the given Hamiltonian. And then we can obtain the expectation of observables associated with the system as a function of time, such as the population$\langle n_j(t) \rangle$, the correlation function $C_{ij}(t)$, etc.

	\section{{The velocity of the correlations}}
\begin{figure}
		\centering \includegraphics[width=0.5\linewidth]{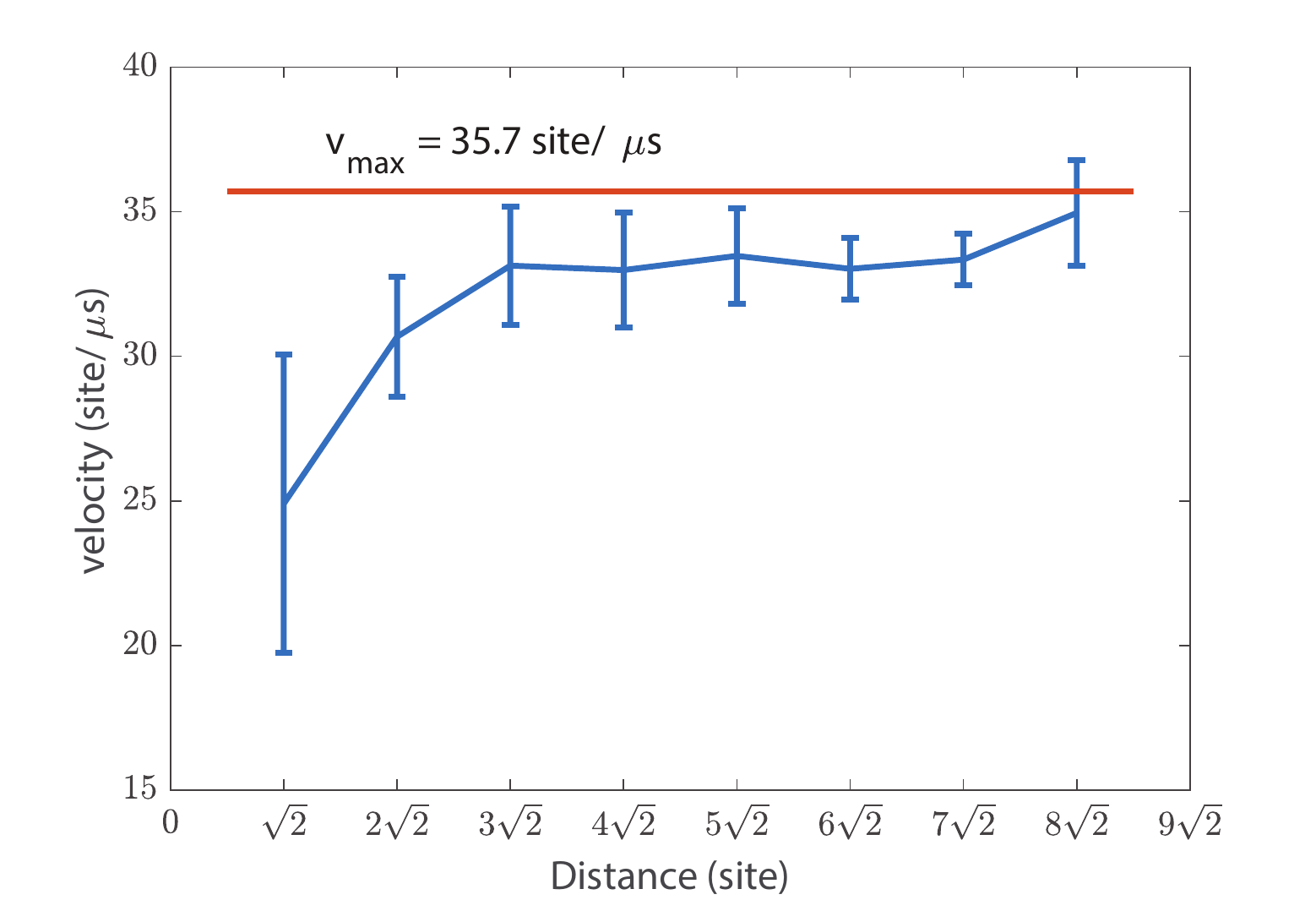}
		\caption{\red{{\textbf{Numerical simulation of the instantaneous velocity as a function of distance in two dimensions.} We illustrate the velocity obtained from a linear fit through the signal positions $d_0\leq d \leq d_0 +3\sqrt{2}$, as a function of distance ($d_0$). It is shown that with the growth of $d_0$, the velocity increase and converges to $v_{max}$ under 1.6MHz random disorder.}
		}}
	\label{figSp}
\end{figure}

	\red{In the experiment in detecting the velocity of the correlations, we used the two-site correlation function between the initial excited qubit U00Q0 and the qubits along the diagonal to determine the propagation velocity in the two dimensional qubit array. As shown in Fig.~2{C} in the main text, we measured the correlation function as a function of time between the initial excitation site and the sites along the diagonal with distance $d=\sqrt{2}$, $d=2\sqrt{2}$, $d=3\sqrt{2}$, and $d=4\sqrt{2}$ sites, respectively. The distance $d$ is defined as the Euclidean distance $\sqrt{\Delta_x^2+\Delta_y^2}$, in which $\Delta_x$($\Delta_y$) is the number of sites between two sites along X(Y) axis. With a Gaussian fittings to the data, we extracted the propagation fronts of the correlations established between the initial excitation and sites at certain distances. Using a linear fitting of experimental fronts with distance, we obtained the velocity of the propagation as $22.2\pm2.0$ site/$\mu s$, which is clearly limited by $v_{max} = 35.7\;  {\rm site}/\mu s$. 
	
	As for the discrepancy between the velocity and $v_{max}$, enlighted by two arguments about the short distance and the disorders in one dimension\cite{burrell2009information,cheneau2012light}, we numerically simulated the correlation function between the initial excitation and each qubit along the diagonal, as a function of time, in a 15$\times$15 qubits array, under reasonable 1.6MHz random disorders. With the correlation function, instantaneous velocity $v(d_0)$ is obtained from a linear fit through the signal positions
    $d_0\leq d \leq d_0 +3\sqrt{2}$. As shown in Fig~\ref{figSp}, it is illustrated that the velocity at short distance and that of long distance are not the same. When $d_0$=$\sqrt{2}$ sites, the instantaneous velocity equals to 24.9 $\pm$ 5.2 site/$\mu s$, which agrees well with the velocity obtained in the experiments (22.2$\pm$ 2.0 site/$\mu s$). With a longer distance, the instantaneous velocity increases. When $d_0$=$8\sqrt{2}$ sites, the instantaneous velocity reached $35.0\pm1.8$ site/$\mu s$. Such results indicate that in a two-dimensional system, the instantaneous velocity under 1.6 MHz random disorder also grows with distance and approaching $v_{max}$ at large distance. Thus, we attribute the reducing of the velocity in our experiments to the short distance and the disorders.   }
	
	\section{{The effect of random disorders and decoherence}}
	
	\begin{figure}[H]
		\centering \includegraphics[width=1\linewidth]{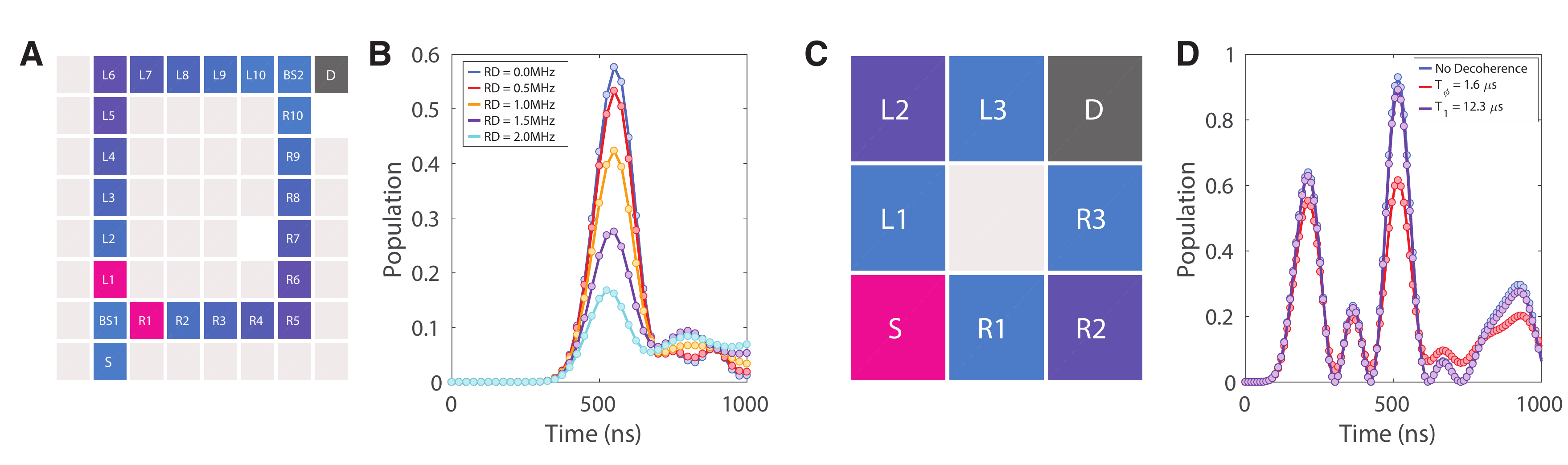}
		\caption{\red{\textbf{Simulating the effect of random disorders and decoherence.}(\textbf{A}) The circuit  diagram  for the simulation of the effect of disorder in single-particle  Mach-Zehnder  interferometer. In the simualtion, we add random disorders (RDs) to both paths to simulation the effect of disorder. The propagation path in the simulation is the same as that of in the Fig.3 of the main text. (\textbf{B}) The numerical simulated population on site $D$ under different RDs. We found that the maximum population decreases with the growth of RDs. (\textbf{C}) The circuit  diagram  for the simulation of the effect of decoherence. In the simulation, the walker is excited in source $S$, goes into two path $\{L\}$ and $\{R\}$ and finally injects into destination $D$. To study the effect of decoherence, We add dephasing or relaxation to every qubit in the array. (\textbf{D}) The numerical simulated population on site $D$ with no decoherence, only dephasing time $T_\phi=1.6\mu$s, and only energy relaxation time $T_1=12.3\mu$s, respectively.}
		}
		\label{figSn}
	\end{figure}
	
	\red{To explain the discrepancy between experimental results and simulated results, we investigated the effect of random disorders and decoherence. In Fig.~\ref{figSn}{A} and {B}, we present the numerical simulated time-resolved population on site $D$ with random disorders. In this simulation, we add random disorders (RDs) on both paths to simulate the effect of disorder {on the final  site $D$}. For each site on both paths, the initial frequency is detuned with $\delta_f$ being a random value between -RD and RD. It is found that with the growth of RD, the maximum population of $D$ decreases accordingly, indicating that the disorder will cause the loss of population in traversing as an effect of reflection.

	As for the effect of decoherence, in Fig.~\ref{figSn}{C} and {D}, we present another numerical simulated results to show the effect of decoherence. We note that we are not able to perform the simulation of the complete multi-qubit system with decoherence because of the extremely large size of the Hilbert space. As a substitute, we simulate a small system containing 8 qubits to investigate the effect of decoherence. The circuit is presented in Fig.~\ref{figSn}{C}. Three conditions are considered, i.e., no decoherence, only energy relaxation time $T_1=12.3\mu$s, and only dephasing time $T_\phi=1.6\mu$s. In Fig.~\ref{figSn}{D}, it is found that when considering the dephasing time $T_\phi$ as $1.6$ $\mu$s, the population of $D$ decays from 0.9 to 0.6 at $t\sim500$ ns in comparing with the other cases. Such result indicates that the dephasing time is also non-negligible, as in our device, the dephasing time is relatively low and comparible to the evolution time. In comparing with the energy relaxation time, the average dephasing time is one order of magnitude lower and in the same order of magnitude as the evolution time, thus will contribute more. 
	}
	\section{Extended data}
	
    \begin{figure*}[htb!]
		\centering \includegraphics[width=0.8\linewidth]{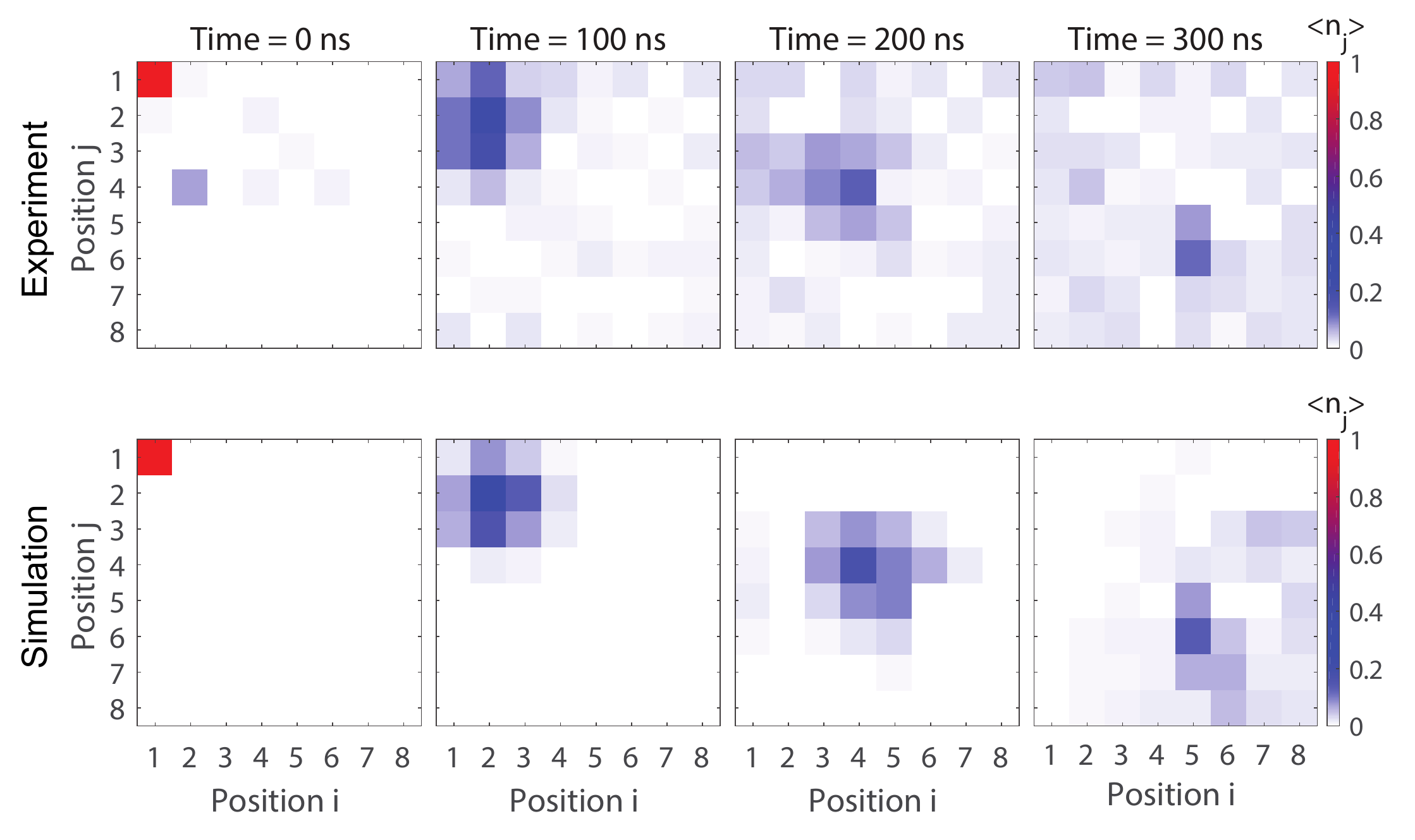}
		\caption{\textbf{Quantum walks in a 2D superconducting qubits array with excitation at U00Q0}. In the state preparation, qubit U00Q0 at the top left corner is excited. Then all qubit are tuned in interaction and finally measured jointly at $t=0$ ns, $100$ ns, $200$ ns, and $300$ ns, respectively. We illustrate the experimental and numerically simulated population $\langle n_{j} \rangle$ of all sites.
		}
		\label{figSd}
	\end{figure*}
	
	\begin{figure*}[htb!]
		\centering \includegraphics[width=0.8\linewidth]{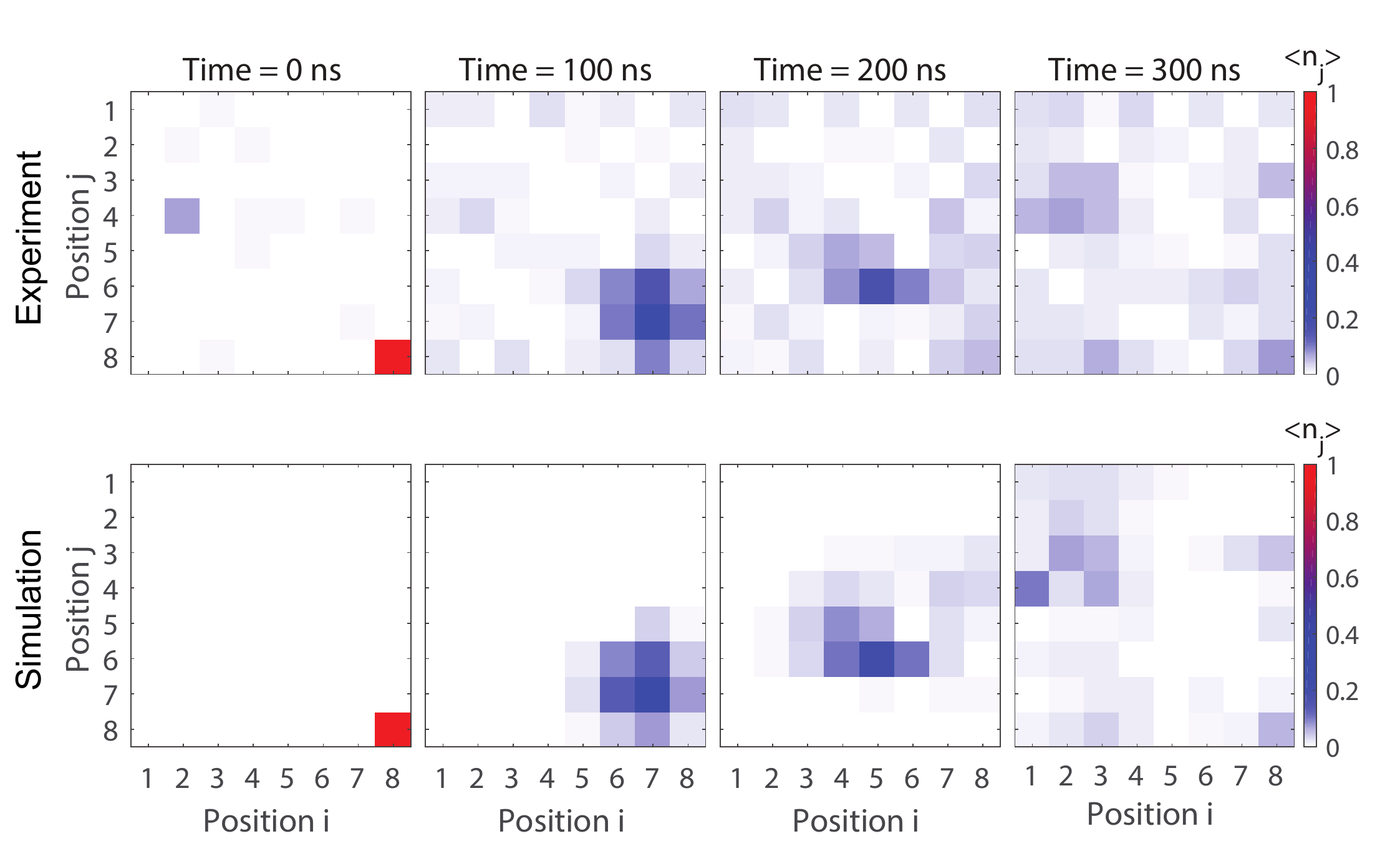}
		\caption{\textbf{Quantum walks in a 2D superconducting qubits array with excitation at U33Q2}. We excite qubit U33Q2 at the bottom right corner in the state preparation and then measure all qubit jointly after an interaction time of $t=0$ ns, $100$ ns, $200$ ns, and $300$ ns, respectively. We illustrate the experimental and numerically simulated population $\langle n_{j} \rangle$ of all sites.}
		\label{figSe}
	\end{figure*}
	
	\begin{figure*}[htb!]
		\centering \includegraphics[width=0.6\linewidth]{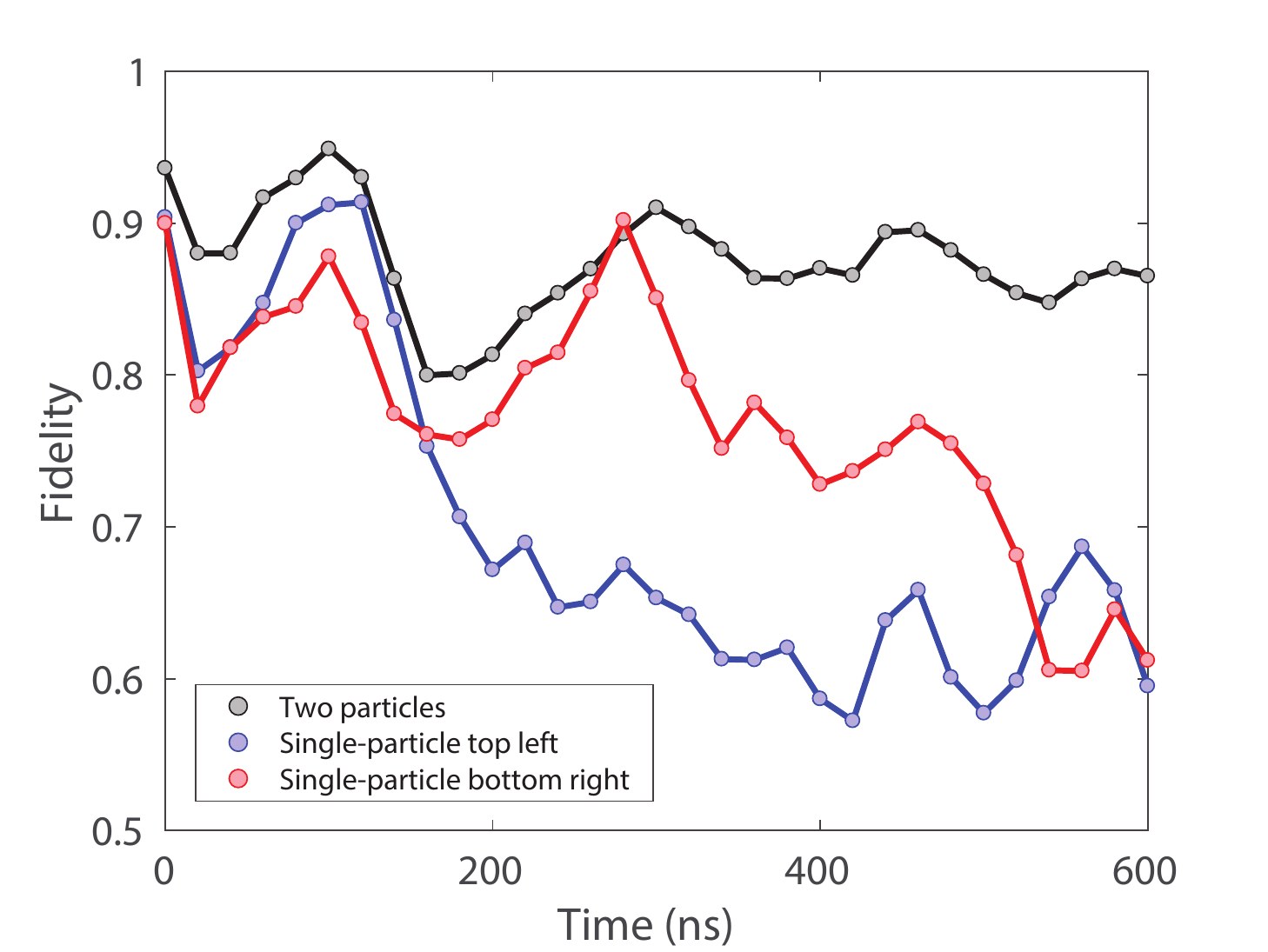}
		\caption{\red{\textbf{Fidelity of the time-resolved evolution of single- and double-particle QWs.} By using the squared statistical overlap as a quantification of fidelity defined by $F=(\sum\sqrt{p_{(i,j)}q_{(i,j)}})^2/\sum p_{i,j}\sum q_{i,j}$, we plot the fidelities as a function of time for double and single particle QWs as shown in Fig.~2, Fig.~\ref{figSd}, and Fig.~\ref{figSe}. Here $p_{(i,j)}$($q_{(i,j)}$) are the experimentally determined (numerically simulated) population distributions.}}
		\label{figSm}
	\end{figure*}
	
	\begin{figure*}[htb!]
		\centering \includegraphics[width=0.5\linewidth]{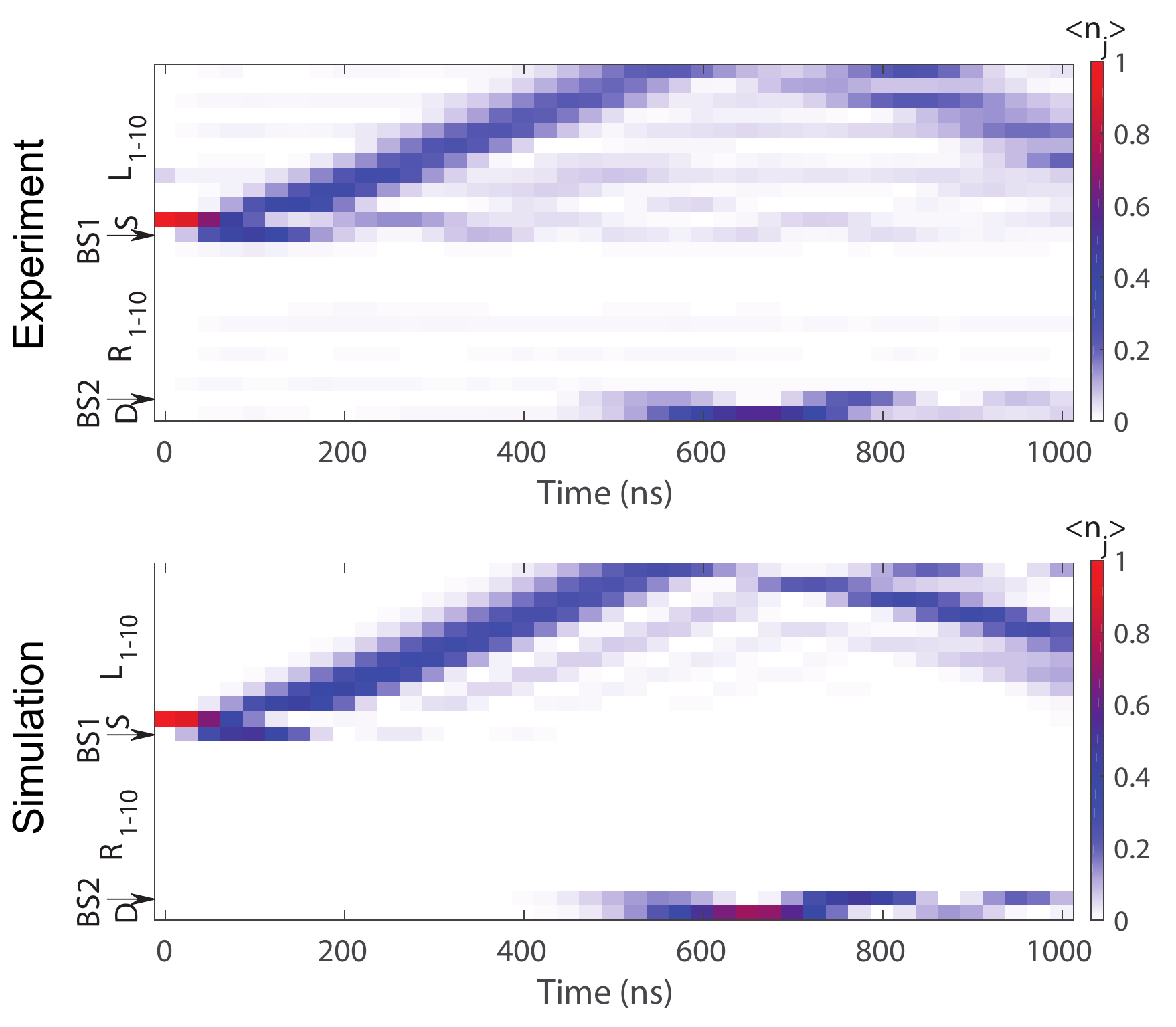}
		\caption{\textbf{The evolution of population $\langle n_{j} \rangle$ in single-particle Mach-Zehnder interferometer with one path blocked.} The circuit diagram is the same as that in Fig. 3{E} in the main text with $\{R\}$ path blocked. The qubit is excited at the source $S$. We illustrate the dynamical evolution of population $\langle n_{j} \rangle$ of all sites in experiment and simulation, respectively.}
		\label{figSf}
	\end{figure*}
	
	
	\begin{figure*}[htb!]
		\centering \includegraphics[width=0.8\linewidth]{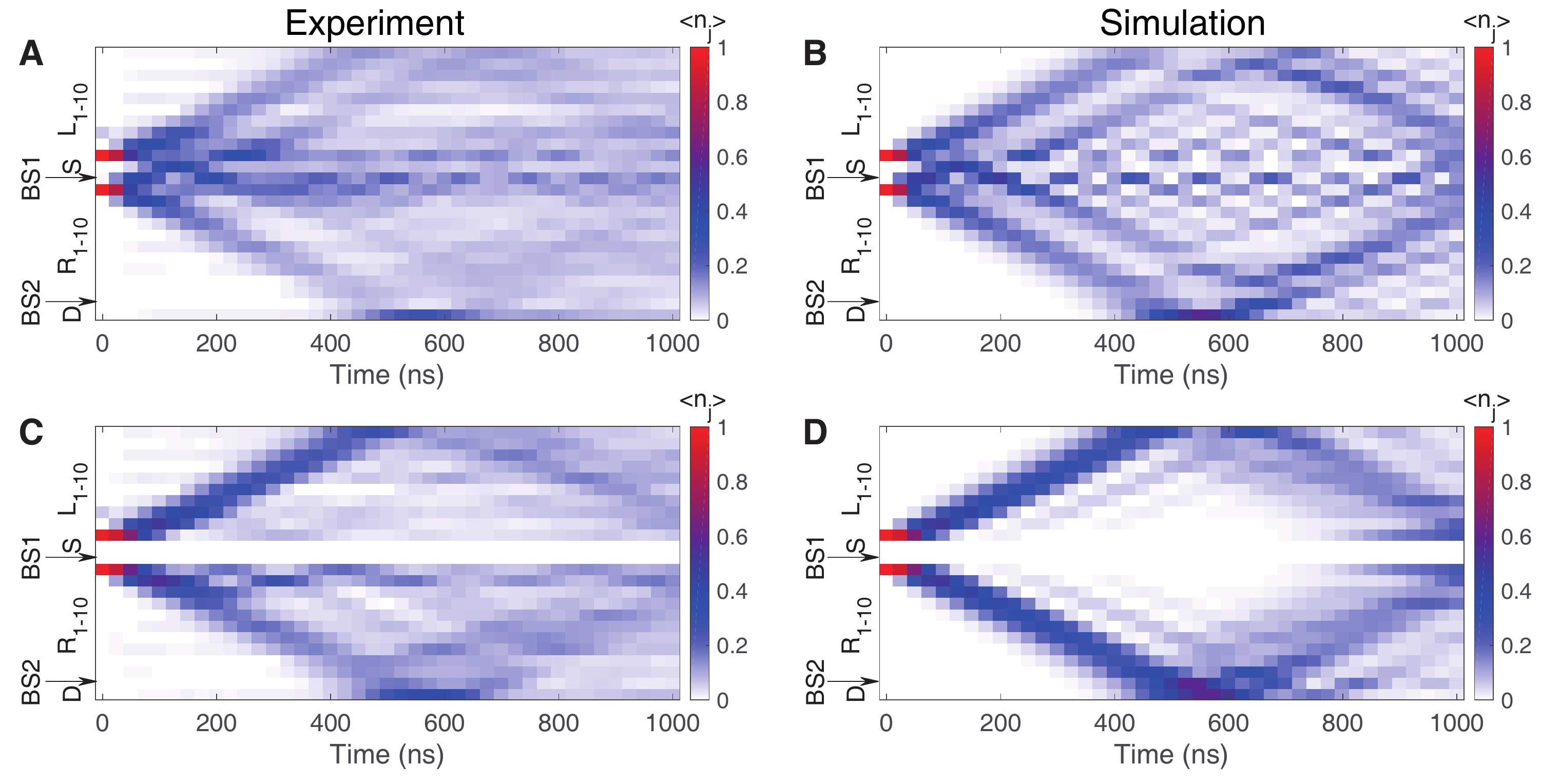}
		\caption{\textbf{The evolution of population $\langle n_{j} \rangle$ with two particles excited at $L_{1}$ and $R_{1}$ in two different situations.} The circuit diagram for (\textbf{A} and \textbf{B}) and (\textbf{C} and \textbf{D}) is the same as that of Fig. 4{A} and 4{C} in the main text, respectively. The difference is in {C} and {D}, sites $BS1$ and $S$ are removed from the interferometer. {B} and {D} are the numerical simulated results with the conditions the same as that of {A} and {C}, respectively.}
		\label{figSg}
	\end{figure*}
	
	\begin{figure}[h]
		\centering \includegraphics[width=0.8\linewidth]{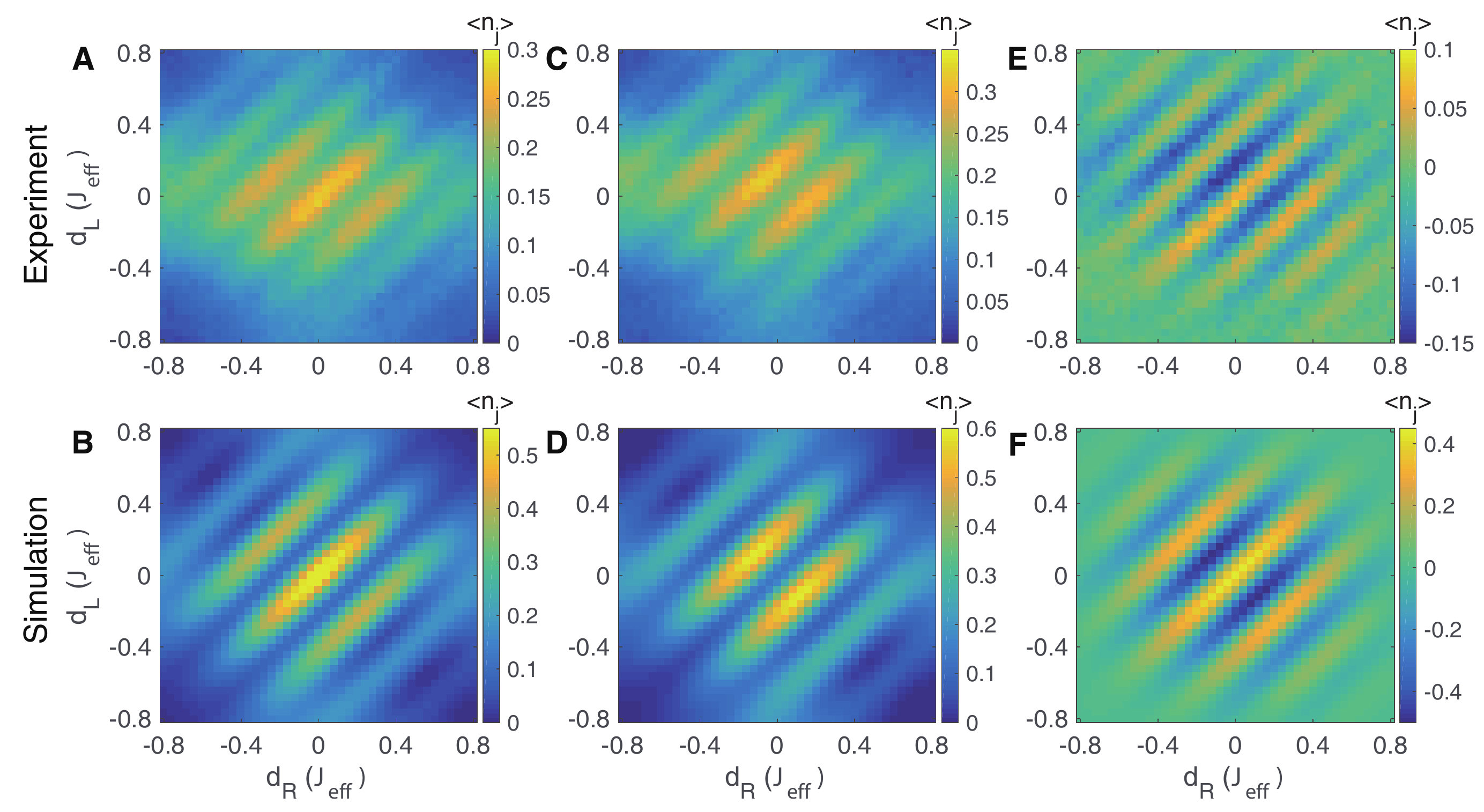}
		\caption{\red{{\textbf{The interaction of two walkers in the Mach-Zehnder interferometer.} In (\textbf{A}) we show the experimental results of Fig.4{B} while (\textbf{C}) depicts the summation of Fig.4{F} and Fig.4{H}. Then in (\textbf{E}) we show the difference between Fig.4{B} and the summation of Fig.4{F} and Fig.4{H}. (\textbf{B}), (\textbf{D}) and (\textbf{F}) show the equivalent simulation results. Both the experimental ({E}) and simulation ({F}) plots show obvious interference fringes, which indicates the interaction between two walkers in the interferometer.}}}
		\label{figSl}
	\end{figure}
	
\renewcommand{\figurename}{Movie.}
\setcounter{figure}{0}
	\begin{figure}
		\caption{\red{(Online only) The time evolution of the populations $\langle n_{j} \rangle$ of all qubits with the two walkers initialized on qubits U00Q0 and U33Q2, in experiment(left) and simulation(right) respectively.}}
		\label{Mov1}
	\end{figure}
	
	\begin{figure}
		\caption{\red{(Online only) The time evolution of the populations $\langle n_{j} \rangle$ of all qubits with the single walker initialized on qubits U00Q0, in experiment(left) and simulation(right) respectively.}}
		\label{Mov2}
	\end{figure}
	
	\begin{figure}
		\caption{\red{(Online only) The dynamical evolution of the populations $\langle n_{j}\rangle$ for single-particle Mach-Zehnder  interferometer with the circuit diagram shown in Fig.~3{A} in the main text.}}
		\label{Mov3}
	\end{figure}
	
	\begin{figure}
		\caption{\red{(Online only) The dynamical evolution of the populations $\langle n_{j}\rangle$ for two-particle Mach-Zehnder interferometer with the circuit diagram shown in Fig.~4{A} in the main text.}}
		\label{Mov4}
	\end{figure}
	
	In Fig.~\ref{figSd} and Fig.~\ref{figSe}, we illustrate the quantum walks of single particle excited at U00Q0 and U33Q2, respectively. \red{The corresponding fidelity of single particle excited at U00Q0 and U33Q2 and fidleity of two particles are shown in Fig.~\ref{figSm}}. The time evolution of population with one path ($\{R\}$) blocked in the interferometer is shown in Fig.~\ref{figSf}. 
	For the case of two particles in the MZ interferometer, the numerical simulated time evolution of all sites are shown in Fig.~\ref{figSg}. \red{In Fig.~\ref{figSl}, we also illustrate the interaction of two walkers in the Mach-Zehnder interferometer} 
	
	
	Moreover, the extended movies (supplied online only) show the time-resolved population  $\langle n_{j} \rangle$ of the two-dimensional quantum walk (Movie \ref{Mov1}, \ref{Mov2}) and the evolution of population $\langle n_{j} \rangle$ in the MZ interferometer with single walker (Movie \ref{Mov3}) and two walkers (Movie \ref{Mov4}). 
	
	\clearpage
 	
	\bibliographystyle{Science}
	
	%
    %
    %
    %
    %